 \documentclass[apj,numberedappendix]{emulateapj}
 \usepackage{epsfig}
 \usepackage{amsmath}
 \usepackage{enumerate}
 \usepackage{natbib}
 \usepackage{hyperref}

\usepackage{color}

\def\simlt{\lower.5ex\hbox{$\; \buildrel < \over \sim \;$}}
\def\simgt{\lower.5ex\hbox{$\; \buildrel > \over \sim \;$}}

\def\sT{\sigma_{\rm T}}

\def\beq{\begin{equation}}
\def\eeq{\end{equation}}
\def\ba{\begin{eqnarray}}
\def\ea{\end{eqnarray}}
\def\bB{{\,\mathbf B}}
\def\bE{{\,\mathbf E}}
\def\bj{{\,\mathbf j}}
\def\bv{{\,\mathbf v}}
\def\Rmax{R_{\rm max}}

\def\BQ{B_Q}

\def\Epar{E_\parallel}

\def\Etarget{E_t}
\def\Ec{\tilde{E}}
\def\mum{\hat{\mu}}
\def\br{{\mathbf r}}

\def\M{{\cal M}}
\def\nmin{n_{\rm min}}
\def\F{{\cal F}}
\def\FF{\bar{\cal F}}
\def\D{{\cal D}}

\def\k{{\mathbf k}}

\def\ee{{\mathbf e}}
\def\epol{{\mathbf e}}
  \def\ang{\vartheta}
   \def\angc{\tilde{\vartheta}}
\def\muc{\tilde{\mu}}
\def\omegac{\tilde{\omega}}
\def\f{f_e}
\def\Pc{\tilde{P}}
  \def\alf{\alpha}
\def\musc{\mu^\prime}
\def\epolsc{{\mathbf e}^\prime}
\def\angsc{\ang^\prime}
\def\Omegasc{\Omega^\prime}
\def\mucsc{{\muc}^\prime}
\def\xisc{\xi^\prime}
\def\sigtot{\sigma_{\rm tot}}
\def\Sect{{\rm Section}} 
\def\Sects{{\rm Sections}} 
  \def\kB{k}
\def\dNsc{\dot{N}_{\rm sc}}
\def\omegares{\omega_{\rm res}}
\def\nn{{\mathbf n}}
\def\gsc{\gamma_{\rm sc}}
\def\bav{\bar{\beta}}
\def\dN{\dot{N}}
\def\pav{\bar{p}}

\def\const{{\rm const}}

\def\lbar{\lambda\llap {--}}
\def\KMC{K_{\rm MC}}
\def\dE{\dot{E}}
\def\Lth{L_{\rm th}}
\def\Iop{I_{\rm LC}}
\def\RLC{R_{\rm LC}}
\def\Fm{\digamma}

\def\Eq{Equation}
\def\Eqs{Equations}

\newbox\grsign \setbox\grsign=\hbox{$>$} \newdimen\grdimen \grdimen=\ht\grsign
\newbox\simlessbox \newbox\simgreatbox \newbox\simpropbox
\setbox\simgreatbox=\hbox{\raise.5ex\hbox{$>$}\llap
     {\lower.5ex\hbox{$\sim$}}}\ht1=\grdimen\dp1=0pt
\setbox\simlessbox=\hbox{\raise.5ex\hbox{$<$}\llap
     {\lower.5ex\hbox{$\sim$}}}\ht2=\grdimen\dp2=0pt
\setbox\simpropbox=\hbox{\raise.5ex\hbox{$\propto$}\llap
     {\lower.5ex\hbox{$\sim$}}}\ht2=\grdimen\dp2=0pt
\def\simgt{\mathrel{\copy\simgreatbox}}
\def\simlt{\mathrel{\copy\simlessbox}}

\begin{document}

\title{Electron-positron flows around magnetars}

\author{Andrei M. Beloborodov}
\affil{Physics Department and Columbia Astrophysics Laboratory,
Columbia University, 538  West 120th Street New York, NY 10027;
amb@phys.columbia.edu}

\begin{abstract}
The twisted magnetospheres of magnetars must sustain a persistent flow of 
electron-positron plasma. The flow dynamics is controlled by the 
radiation field around the hot neutron star. The problem of plasma motion in 
the self-consistent radiation field is solved using the method of virtual beams.
The plasma and radiation exchange momentum via resonant 
scattering and self-organize into the ``radiatively locked'' outflow
with a well-defined, decreasing Lorentz factor. 
There is an extended zone around the magnetar where the plasma flow 
is ultra-relativistic; its Lorentz factor is self-regulated so that it 
can marginally scatter thermal photons. The flow becomes slow and opaque 
in an outer equatorial zone, where the decelerated plasma accumulates and 
annihilates; this region serves as a reflector for the thermal photons emitted 
by the neutron star. The $e^\pm$ flow carries electric current, which is 
sustained by a moderate induced electric field. 
The electric field maintains a separation between the electron and positron 
velocities, against the will of the radiation field.
The two-stream instability is then inevitable,
and the induced turbulence can generate low-frequency emission. In particular,
radio emission may escape around the magnetic dipole axis of the star. 
Most of the flow energy is converted to hard X-ray emission, which is examined
in the accompanying paper.
\end{abstract}



\section{Introduction}

The observed activity of magnetars is believed to be caused by their surface 
motions, which are driven by strong internal stresses. 
The magnetosphere is anchored in the neutron
star and twisted by the surface motions, resembling the 
behavior of the solar corona. As a result it becomes 
non-potential, $\nabla\times\bB\neq 0$, and threaded by electric currents 
(Thompson et al. 2002; Beloborodov \& Thompson 2007). 
The currents flow along $\bB$, i.e., the 
twisted magnetosphere remains nearly force-free, $\bj\times\bB=0$.
Numerical models of dynamic twisted magnetospheres of magnetars 
(Parfrey et al. 2012, 2013) show how the twist creates spindown 
anomalies and initiates giant flares when the magnetosphere 
is ``overtwisted'' and loses equilibrium.

Free energy stored in the magnetospheric twist is gradually dissipated through 
the continual electron-positron discharge that sustains the electric current.
As a result, the magnetosphere tends to ``untwist'' 
on the characteristic ohmic timescale of a few years.
Electrodynamics of untwisting is quite peculiar: 
whenever the crustal motions stop or slow down,
the electric currents tend to be quickly removed from the magnetospheric field 
lines with small apex radii $\Rmax$  (Beloborodov 2009).
Currents have longest lifetimes on field lines with 
$\Rmax\gg R$ (where $R\approx 10$~km is the star radius) and
form an extended ``j-bundle'' (Figure~1). The j-bundle has 
a sharp boundary, which gradually shrinks toward the magnetic dipole axis.
In particular, the footprint of the j-bundle on the neutron star surface, which 
may be observed as a hot spot, shrinks with time.
Shrinking hot spots were indeed reported in ``transient magnetars'' whose 
magnetospheres were temporarily activated and then gradually relaxed back to 
the quiescent state (see Figure~5 in Beloborodov 2011a and refs. therein).

\begin{figure*}[t]
\epsscale{1.03}
\plotone{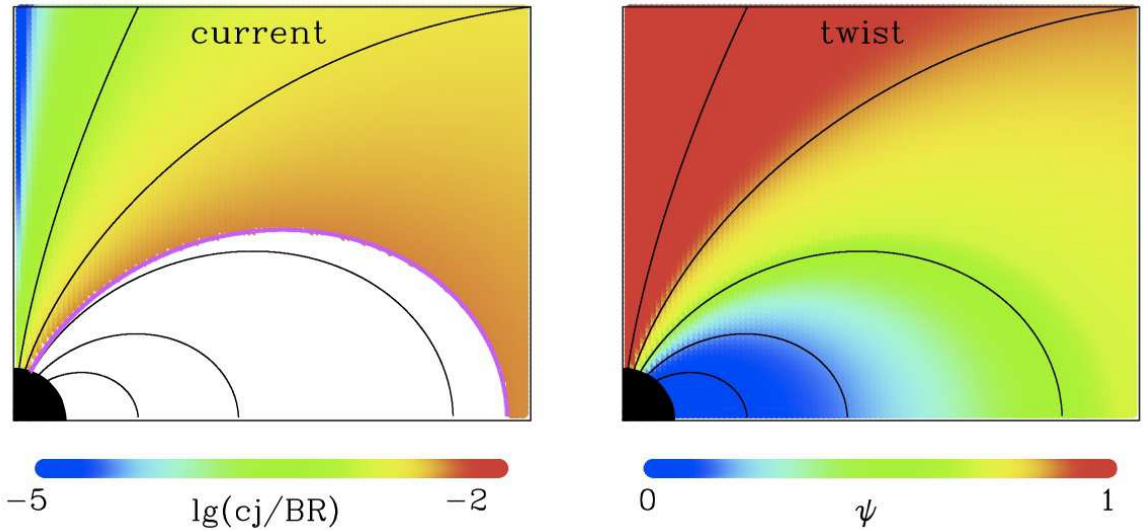}
\caption{
Snapshot of a slowly untwisting magnetosphere. 
In this example, a global twist with a uniform amplitude $\psi=0.2$ 
was implanted into the dipole magnetosphere at $t=0$, and 
the snapshot shows the magnetosphere at $t\sim 1$~yr. 
Details of the calculations are described in Beloborodov (2009). 
The plane of the figure is the poloidal cross section of the magnetosphere.
The black curves are the poloidal magnetic field lines. The magnetosphere 
is symmetric about the vertical axis and the equatorial plane; therefore,
the figure only shows one quarter of the poloidal cross section. 
The neutron star is shown by the black circle (radius $R\approx 10$~km).
Left panel: Current density $j$ normalized to $BR/c$. 
The region from which currents have 
been pulled into the star (the potential ``cavity'' with $j=0$) is shown 
in white. The boundary between the cavity and the j-bundle 
(magenta curve) expands with time, i.e. the j-bundle 
shrinks toward the vertical axis. 
Right panel: Twist amplitude $\psi$ at the same time. 
The twist amplitude is defined for each closed field line as the 
azimuthal displacement of its footpoint in the southern hemisphere
relative to the footpoint in the northern hemisphere.
}
\end{figure*}

The j-bundle must be filled with relativistic plasma that carries the electric 
current $\bj=(4\pi/c)\nabla\times\bB$. The plasma is continually created by 
$e^\pm$ discharge near the star and must expand along the extended field 
lines. The plasma emits persistent nonthermal emission, converting the 
dissipated twist energy to radiation. In the accompanying paper 
(Beloborodov 2013), we calculate the spectrum of produced 
radiation and show that it forms the observed hard X-ray component with a 
peak around 1~MeV. The present paper studies in more detail the dynamics 
of the $e^\pm$ plasma.

Previously, semi-transparent plasma flows around magnetars were invoked to 
explain the deviation of the observed 1-10~keV emission from a thermal spectrum
(Thompson et al. 2002).
It was usually assumed that the magnetar corona is filled with positive 
and negative charges that are counter-streaming with mildly relativistic 
speeds. The counter-streaming picture was motivated by the fact that electric 
current $\bj$ must flow along the twisted magnetic field lines. The coronal 
plasma must be nearly neutral; it can easily carry the required 
current if the opposite charges with densities $n_+=n_-$ flow in the 
opposite directions, so that $j=e(v_+n_+-v_-n_-)$ where $v_+v_-<0$. 
The velocities $v_\pm$ are free parameters in this phenomenological model, 
which may be adjusted so that resonant scattering of thermal X-rays in the corona 
reproduces  the 1-10~keV part of the magnetar spectrum 
(e.g. Fern\'andez \& Thompson 2007; Nobili et al. 2008; Rea et al. 2008). 

The counter-streaming model has, however, a problem. Note that the thermal 
radiation of the magnetar ($h\nu\sim 1$~keV) is resonantly scattered at large 
radii $r\sim 10R\approx 100$~km where $B\sim 10^{11}-10^{12}$~G.\footnote{
       This assumes that photons are scattered by electrons, not ions.} 
In this region, the plasma is strongly pushed by radiation away from the 
star and the counter-streaming model needs an electric field $\Epar$ that 
forces charges of the right sign to move toward the star against the 
radiative drag. At the same time, the electric field acts on the 
opposite charges and accelerates them away from the star, 
cooperating with the radiative push. 
In the presence of $e^\pm$ plasma (which is inevitably created near 
magnetars), the outward acceleration generates relativistic
particles, and no self-consistent solution exists for the 
mildly-relativistic counter-streaming model.

In this paper (and the accompanying paper Beloborodov (2013)),  
we develop a different picture of plasma circulation in the magnetar corona. 
It is schematically shown in Figure~2. 
The outer corona is inevitably filled with $e^\pm$ pair plasma of a high density 
$n$, which is larger than $j/ec$ by the ``multiplicity factor'' $\M\gg 1$; 
in this respect it resembles the flow along the open field lines 
of a rapidly rotating, strongly magnetized neutron star
(e.g. Hibschman \& Arons 2001; Thompson 2008a; Medin \& Lai 2010).
Pairs are created in the ``adiabatic zone'' $B>10^{13}$~G where the flow 
energy is reprocessed into particles with Lorentz factors $\gamma\sim 20$ 
(Beloborodov 2013); their multiplicity $\M\sim 10^2$ is basically set by energy 
conservation and mainly controlled by the discharge voltage.
Both electrons and positrons outflow from the magnetar, and radiation pressure 
forces the particles to accumulate  in the equatorial plane of the magnetic dipole, 
where they annihilate. The required current $\bj=(c/4\pi)\nabla\times\bB$ is 
sustained in the outflow by a moderate electric field $\Epar$. 
This field is self-consistently generated to 
maintain a small difference between the velocities of the $\pm$ charges,
$(v_+-v_-)/v_\pm\sim \M^{-1}\ll 1$, so that the condition $e(n_+v_+-n_-v_-)=j$ 
is satisfied with $n_+\approx n_-$ and $v_+v_->0$. In the simplest, two-fluid
model (\Sect~3) the velocities $v_\pm$ tend 
to be ``locked'' by the balance of two forces, electric and radiative.

The coronal outflow significantly changes the radiation it interacts with via 
scattering. The problem of outflow dynamics can be formulated as a problem of 
self-consistent radiative transfer where particles and photons exchange energy 
and momentum as they flow away from the neutron star. This problem is solved
in this paper using a specially designed numerical method.

The paper is organized as follows. \Sect~2 discusses the creation of $e^\pm$ 
pairs and their circulation in the inner and outer magnetosphere. \Sect~3 presents 
the model of a radiatively locked outflow in its simplest version using a two-fluid 
description and assuming an optically thin magnetosphere. \Sect~4 discusses 
the two-stream instability in the $e^\pm$ flow and the origin of low-frequency 
emission from magnetars.
Then, in \Sects~5 and 6 we formulate and solve the full problem where an 
outflow with a broad momentum distribution function and significant optical 
depth interacts with the neutron-star radiation. The numerical method and 
results are described in \Sect~6. Our conclusions are summarized in \Sect~7.


\section{Creation and circulation of $e^\pm$ pairs}

The creation of $e^\pm$ pairs by an accelerated particle is a two-step process:
the particle generates a high-energy photon (via resonance scattering) and then 
the photon converts to an $e^\pm$ pair in the strong magnetic field. 
An accelerated electron (or positron) can resonantly scatter photons of energy 
$\hbar\omega$ once it reaches the Lorentz factor required by the resonance
condition $\gamma(1-\beta\cos\ang)\omega=\omega_B$, where
$\ang$ is the angle between the photon and the electron velocity and 
$\omega_B=eB/m_ec$. When $\ang$ is not small, the resonance condition gives
\beq
\label{eq:res1}
  \gsc\sim 10^3 B_{14}\left(\frac{\hbar\omega}{1 \rm ~keV}\right)^{-1}.
\eeq
The scattered photons are boosted in energy by the factor of $\sim \gsc^2$.
Such high-energy photons quickly convert to $e^\pm$ pairs in the strong magnetic 
field, creating more particles near the star.
A similar process of $e^\pm$ creation operates in the polar-cap discharge 
of ordinary pulsars, but in a different mode. In ordinary pulsars, the 
high-energy photons convert to $e^\pm$ with a significant delay. The scattered
photon initially moves nearly parallel to $\bB$ and converts to $e^\pm$ 
only when it propagates a sufficient distance where its angle 
$\theta_\gamma$ with respect to $\bB$ increases so that the threshold 
condition for conversion is satisfied. This delay leads to the large 
unscreened voltage in pulsar models.

In contrast, the magnetic field of magnetars is so strong that pair 
creation can occur immediately following resonant scattering
(Beloborodov \& Thompson 2007). 
The energy of a resonantly scattered photon is related to its emission 
angle $\theta_\gamma$ by
\beq
  E(\theta_\gamma)=\frac{E_B}{\sin^2\theta_\gamma}
   \left[ 1- \left(\cos^2\theta_\gamma
       +\frac{m_e^2c^4}{E_B^2}\sin^2\theta_\gamma\right)^{1/2}\right],
\eeq
where $E_B=(2B/\BQ+1)^{1/2} m_ec^2$ is the energy of the first excited 
Landau level and $\BQ=m_e^2c^3/\hbar e\approx 4.4\times 10^{13}$~G. 
The scattered photon may immediately be above the threshold for conversion,
$E>E_{\rm thr}=2m_ec^2/\sin\theta_\gamma$, if $B>4\BQ$.
Therefore, $e^\pm$ discharge in magnetars can screen $\Epar$ more 
efficiently than in ordinary pulsars and buffer the voltage growth once 
the Lorentz factors of accelerated particles reach $\gsc$ given in \Eq~(\ref{eq:res1}).

\subsection{Pair creation on field lines with apexes $\Rmax\simlt 2R$}

The discharge on twisted closed field lines can be explored using a 
direct numerical experiment where plasma is represented by a large 
number of individual particles in the self-consistent electric field.
The existing numerical simulations (Beloborodov \& Thompson 2007) 
describe the discharge on field lines that extend to a moderate radius
$\Rmax\simlt 2R$, where $R$ is the radius of the neutron star.
The magnetic field is ultrastrong everywhere along such field lines,
$B\gg\BQ$, and resonant scattering events may effectively be treated 
as events of pair creation --- a significant fraction of scattered
photons immediately convert to $e^\pm$.

The simulations demonstrate that voltage and pair creation self-organize 
in the twisted magnetosphere so that a particle on average scatters 
$\sim 1$ photon as it travels through the electric circuit, maintaining 
the near-critical multiplicity of pair creation $\M\sim 1$. 
This criticality condition regulates the induced voltage to $\Phi_e\sim 10^9$~V, 
which accelerates $e^\pm$ particles to Lorentz factors $\gamma\sim 10^3$.
The electric circuit operates as a {\it global} discharge, in the sense 
that the accelerating voltage is distributed along the entire field line 
between its footpoints on the star. It is quite different from the 
localized ``gap'' that is usually considered above polar caps in pulsars.

The discharge fluctuates on the light-crossing timescale $\sim R/c$ and 
persists in the state of self-organized criticality. The behavior of the global
circuit resembles a continually repeating lightning: voltage between the 
footpoints of the field line quasi-periodically builds up and discharges 
through the enhanced production of charges. The average plasma density in the 
circuit $n$ is close to the minimum density $\nmin=j/ec$, as required by 
the criticality condition $\M=n/\nmin \sim 1$. 

The global-discharge picture applies only to field lines with 
$\Rmax\simlt 2R$ and becomes irrelevant when the currents are erased 
in the inner magnetosphere as shown in Figure~1.
Then the observed activity must be associated with currents on field lines
with large $\Rmax$, i.e. extending far from the star. 

\begin{figure}
\epsscale{1.03}
\plotone{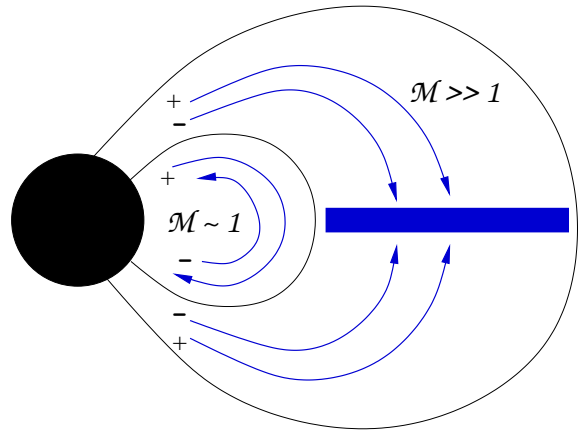}
\medskip
\caption{
Schematic picture of plasma circulation in the magnetosphere with
surface $B\sim 10^{15}$~G.  Two regions are indicated. (1) ``Inner corona.''
Here $e^\pm$ have a moderate multiplicity $\M\sim 1$. The particles do
not stop in the equatorial plane.  The electric field $\Epar$ ensures that
electrons and positrons circulate in the opposite directions
along the magnetic field lines, maintaining the electric current
demanded by $\nabla\times\bB$. The particles are lost as they reach the
footpoints of the field line and continually replenished by pair creation.
(2) ``Outer corona'' --- extended field lines with $\Rmax\gg R$.
  Electrons and positrons are created by the discharge near the star and
  some of them flow outward to the region of weaker $B$.
  Here resonant scattering
  enhances the pair multiplicity, $\M\gg 1$, and decelerates the outflow.
  The $e^\pm$ particles stop at the apexes of magnetic field lines (blue
  region in the equatorial plane), accumulate, and annihilate there.
  The number fluxes of electrons and positrons toward the annihilation
  region differ by a small fraction $\sim\M^{-1}$, so that the outflow
  carries the required electric current $\bj=(c/4\pi)\nabla\times\bB$.
  Electrodynamics of the twist dissipation implies that the inner corona 
  is less likely to be active, as the electric currents are erased by the expanding
  cavity (Figure~1); the observed activity tends to concentrate on extended 
  field lines that form the outer corona.
}
\end{figure}

\subsection{Pair creation on field lines with apexes $\Rmax\gg R$}

The discharge on extended field lines may be expected to have a similar 
threshold voltage $\Phi_e\sim 10^9$~V, because the conversion of upscattered 
photons to $e^\pm$ is efficient near the field-line footpoints where $B\gg\BQ$.
In this zone, particles are able to resonantly scatter soft X-rays once 
they are accelerated to $\gamma\sim 10^3$ (\Eq~\ref{eq:res1}), 
which requires $\Phi_e\sim 10^9$~V. Further growth of voltage 
would cause excessive creation of $e^\pm$ moving in 
both directions, toward and away from the star, leading to efficient 
screening of $\Epar$.
  
A large fraction of the created particles must outflow to $r\gg R$ along the 
extended field lines. The rate of 
resonant scattering by a relativistic particle {\it increases} as it moves 
from $B\gg \BQ$ to $B\simlt \BQ$. The particle scatters 
many more photons, because the resonance condition shifts toward photons 
of lower energy $\hbar\omega_{\rm res}\propto B$ whose number density is 
larger. Note also that the effective cross section for resonant scattering
$\sigma_{\rm res}=2\pi^2 r_ec/\omega_{\rm res}$ increases as $B^{-1}$
(here $r_e=e^2/m_ec^2$ is the classical electron radius).
Practically all photons scattered by the
outflowing particles in the region $B\simgt 10^{13}$~G convert to $e^\pm$;
detailed Monte-Carlo simulations of this process are presented in the 
accompanying paper (Beloborodov 2013).
In essence, the particles outflowing from the discharge zone lose energy 
to photon scattering, 
and this energy is transformed to new generations of $e^\pm$.  As a result, 
the $e^\pm$ multiplicity of the outflow increases from $\M\sim 1$ to 
$\M\sim 100$.
This implies that there is no charge starvation in the outer corona --- there are 
plenty of charges to conduct the current demanded by the twisted magnetic field. 

Pair creation  sharply ends near the surface of $B\approx 10^{13}$~G; outside 
this surface the resonantly scattered photons are not absorbed.
The steady relativistic outflow without pair creation 
maintains $\M=const$ along the magnetic field lines. This follows from 
conservation of magnetic flux, charge, and particle number, which give
$n/B=const$, $j/B=const$, and $n/j=const$ along the field line.

\subsection{Global circulation of pair plasma}

The picture of plasma circulation in the magnetar magnetosphere is 
summarized in Figure~2. 
The inner corona (field lines with $\Rmax$ of a few neutron-star
radii) is filled with the ultra-relativistic counter-streaming $e^-$ and 
$e^+$. In this region, resonant scattering is marginally efficient 
and a global discharge operates as described in \Sect~2.1, with 
multiplicity $\M\sim 1$. Outside this region, pair multiplicity is 
much higher and the electric current is organized with
both $e^-$ and $e^+$ outflowing from the star.
The exact location of the boundary between the two regions depends 
on the strength of the magnetic field of the star.

In the outer corona, the opposite flows in the nothern and southern 
hemispheres meet in the equatorial plane of the magnetic dipole and 
stop there. Two effects prevent their inter-penetration.
(1) Radiative drag is strong in the outer corona and pushes both 
nothern and southern flows toward the equatorial plane (see \Sect~3 
below). 
(2) When the two opposite flows try to penetrate each other, 
a two-stream instability develops. As a result, a strong Langmuir 
turbulence is generated, which inhibits the penetration.
This effect is particularly important in the transition region between 
the outer and inner corona. For these field lines, resonant scattering 
is efficient enough to generate $\M>1$ but not strong enough to stop 
the pair plasma in the equatorial plane. Then the colliding northern and 
southern flows are stopped by the two-stream instability. This behavior 
contrasts with the inner corona where the induced electric field enforces the 
counter-streaming of $e^-$ and $e^+$, i.e. the opposite flows with $\M\sim 1$ 
are forced to penetrate each other despite the two-stream instability.
  
The density of $e^\pm$ pairs accumulated in the outer equatorial 
region (shown in blue in Fig.~2) is regulated by the annihilation
balance. In a steady state, 
the annihilation rate is given by $\dot{N}_{\rm ann}\approx 2\M (I/e)$,
where $I$ is the electric current through the annihilation region.
The corresponding annihilation luminosity is $L_{\rm ann}=2m_ec^2\M\, I/e$.


\medskip

\section{Radiatively locked coronal flow}

Dynamics of the $e^\pm$ flow in the outer corona ($r\gg R$) is 
influenced by resonant scattering, which exerts a strong force 
$\F$ on the particles along the magnetic field lines. 
$\F$ vanishes only if the particle has the ``saturation momentum'' 
$p_\star$ such that the radiation 
flux measured in the rest frame of the particle is perpendicular to $\bB$. 
In the simplest case of
a weakly twisted dipole magnetosphere  exposed 
to central radiation $p_\star$ is given by (Appendix~B),\footnote{
    This expression is valid in the region where $1-B_r/B>(R/r)^2$. 
    In this region, stellar radiation may be approximated as a central 
    flow of photons, neglecting the angular size of the star $\sim R/r$.}
\beq
\label{eq:sat1}
    p_\star(r,\theta)=\frac{2\cos\theta}{\sin\theta},
\eeq 
where momentum is in units of $m_ec$.
The radiative force always pushes the particle toward $p=p_\star$. 
The strength of this effect
may be measured by the dimensionless ``drag coefficient,'' 
\beq
\label{eq:drag}
   \D\equiv \frac{r\F}{p\,m_ec^2}.
\eeq
Momentum $p_\star$ is a strong attractor in the sense that deviations 
$p-p_\star$ generate $\D\gg 1$ in the outer corona (see Appendix~A).

Even an extremely strong radiative drag does not imply that $e^+$ and $e^-$ 
acquire exactly equal velocities 
$\beta_+=\beta_-=\beta_\star=p_\star(1+p_\star^2)^{-1/2}$. 
Such a ``single-fluid'' flow would be unable to carry the required electric 
current $j$, and $\Epar$ must be induced to ensure a sufficient 
velocity separation $\beta_+-\beta_-$.
Below we describe the two-fluid model with $\beta_+\neq\beta_-$.

\subsection{Two-fluid model: basic equations}
\label{sec:fluid}

Consider an $e^\pm$ flow in the region where no new pairs are produced. 
 In a steady state, the fluxes of electrons and positrons are conserved 
  $\nabla\cdot(n_\pm\bv_\pm)=0$. This implies $\nabla\cdot\bj_\pm=0$
  where $\bj_\pm=\pm e n_\pm v_\pm$ are the contributions of electrons
  and positrons to the net electric current $\bj=\bj_++\bj_-$.  
  Since $e^\pm$ move along the magnetic field lines, 
  $\bj_\pm=\alpha_\pm\bB$ where $\alpha_\pm$ are scalar functions.
  From $\nabla\cdot\bj_\pm=0$ and $\nabla\cdot\bB=0$ one gets  
  $\bB\cdot\nabla\alpha_\pm=0$. Therefore,
\beq
\label{eq:jpm}
   \frac{j_+}{B}=const, \qquad \frac{j_-}{B}=const,
\eeq
are constant along the field line.
The net current $j=j_++j_-$ is fixed by the condition
$j=(c/4\pi)|\nabla\times\bB|$. The multiplicity of pairs is defined by
\beq
   \M=\frac{j_++|j_-|}{j}.
\eeq
Here ``$+$'' corresponds to positrons, which carry current $j_+>0$
and ``$-$'' corresponds to electrons, which carry $j_-<0$;
we assume a positive net current $j=j_++j_-$ for definiteness.
A counter-streaming model ($v_+v_-<0$) would have $\M=1$. 
A charge-separated outflow ($n_-=0$) would also have $\M=1$. 
We study here pair-rich outflows with $\M>1$.
                                                                                
The outflowing $e^\pm$ plasma must be nearly neutral,\footnote{
     Here we neglect rotation of the neutron star and its magnetosphere,
     which is a good approximation everywhere except the open field-line bundle 
     that connects the star to the light cylinder. 
     In a more exact model, the Gauss law in the co-rotating frame 
     $\nabla\cdot\bE=4\pi(\rho+\rho_v)$ includes the effective vacuum charge 
     density $\rho_v={\mathbf \Omega}\cdot\bB/2\pi c$ where ${\mathbf\Omega}$ 
     is the angular velocity of the star (Goldreich \& Julian 1969). 
     Then the neutrality condition becomes $e(n_+-n_-)+\rho_v=0$.
     Magnetars rotate slowly (typical $\Omega\sim 1$~rad/s) and hence a small
     fraction of their magnetic flux is open, typically $\simlt 10^{-4}$.
     In the main, closed magnetosphere, 
     the condition $|\rho_v/e|\ll |j/ec|<n_\pm$ is satisfied,
     and neutrality requires $n_+\approx n_-$ with a high accuracy.}
\beq
   n_+\approx n_-,
\eeq
otherwise a huge electric field would be generated that would restore
neutrality. Using this condition, one finds that $en_+v_+-en_-v_-=j$ 
is satisfied if
\beq
\label{eq:j}
    1-\frac{\beta_-}{\beta_+}=\frac{2}{\M+1}.
\eeq
  A deviation from this condition implies a mismatch between the 
  conduction current $j_++j_-$ and $(c/4\pi)\nabla\times\bB$, which would 
  induce a growing electric field according to Maxwell equation 
  $\partial\bE/\partial t=c\nabla\times\bB-4\pi\bj$.
An electric field $\Epar$ must be established in the outflow to 
sustain the condition~(\ref{eq:j}) against the radiative drag that tends 
to equalize $\beta_-$ and $\beta_+$ at $\beta_\star$. This moderate electric 
field must be self-consistently generated by a small deviation from neutrality, 
$\delta n=n_+-n_-\ll n_\pm$.

The two-fluid dynamics of the outflow is governed by two equations,
\begin{eqnarray}
\label{eq:dyn}
    m_ec^2\frac{d\gamma_\pm}{dl}=\F(\gamma_\pm)\pm e\Epar, 
\end{eqnarray}
where $l$ is length measured along the magnetic field line.
In the region of strong drag, $|\D|\gg 1$, the left-hand side is small compared with 
$\F$; then the radiative and electric 
forces on the right-hand side nearly balance each other,
$\F(\gamma_+)\approx-e\Epar$ and $\F(\gamma_-)\approx e\Epar$. This implies,
\beq
\label{eq:bal}
   \F(\gamma_+)\approx -\F(\gamma_-) \qquad (|\D|\gg 1).
\eeq
Equations~(\ref{eq:j}) and (\ref{eq:bal}) describe the
``radiatively locked'' two-fluid current with pair multiplicity $\M$.
The solution $\gamma_\pm$ to these equations exists if $\M>1$.
Note that $\beta_-<\beta_\star<\beta_+$ in the radiatively-locked state.
                                                     
Let us now relax the assumption $|\D|\gg 1$.
Then the inertial term $m_ec^2d\gamma_\pm/dl$ must be retained
in \Eqs~(\ref{eq:dyn}), i.e. we now deal with differential equations for 
$\gamma_\pm$.
Since $\gamma_-$ and $\gamma_+$ are not independent --- they are related by 
condition~(\ref{eq:j}) --- it is sufficient to solve one dynamic equation, 
e.g. for $\gamma_+$ (and use the dynamic equation for $\gamma_-$ to exclude 
$\Epar$). Straightforward algebra gives,
\begin{eqnarray}
\label{eq:gam_2F}
  m_ec^2\frac{d\gamma_+}{dl} &=& \frac{\F(\gamma_+)+\F(\gamma_-)}
       {1+d\gamma_-/d\gamma_+},  \\
   \frac{d\gamma_-}{d\gamma_+} &=& \left(\frac{\M-1}{\M+1}\right)^2
               \left(\frac{\gamma_-}{\gamma_+}\right)^3.
\end{eqnarray}

\begin{figure*}[t]
\begin{tabular}{cc}
\includegraphics[trim = 0cm 0cm 3cm 0cm, width=0.42\textwidth]{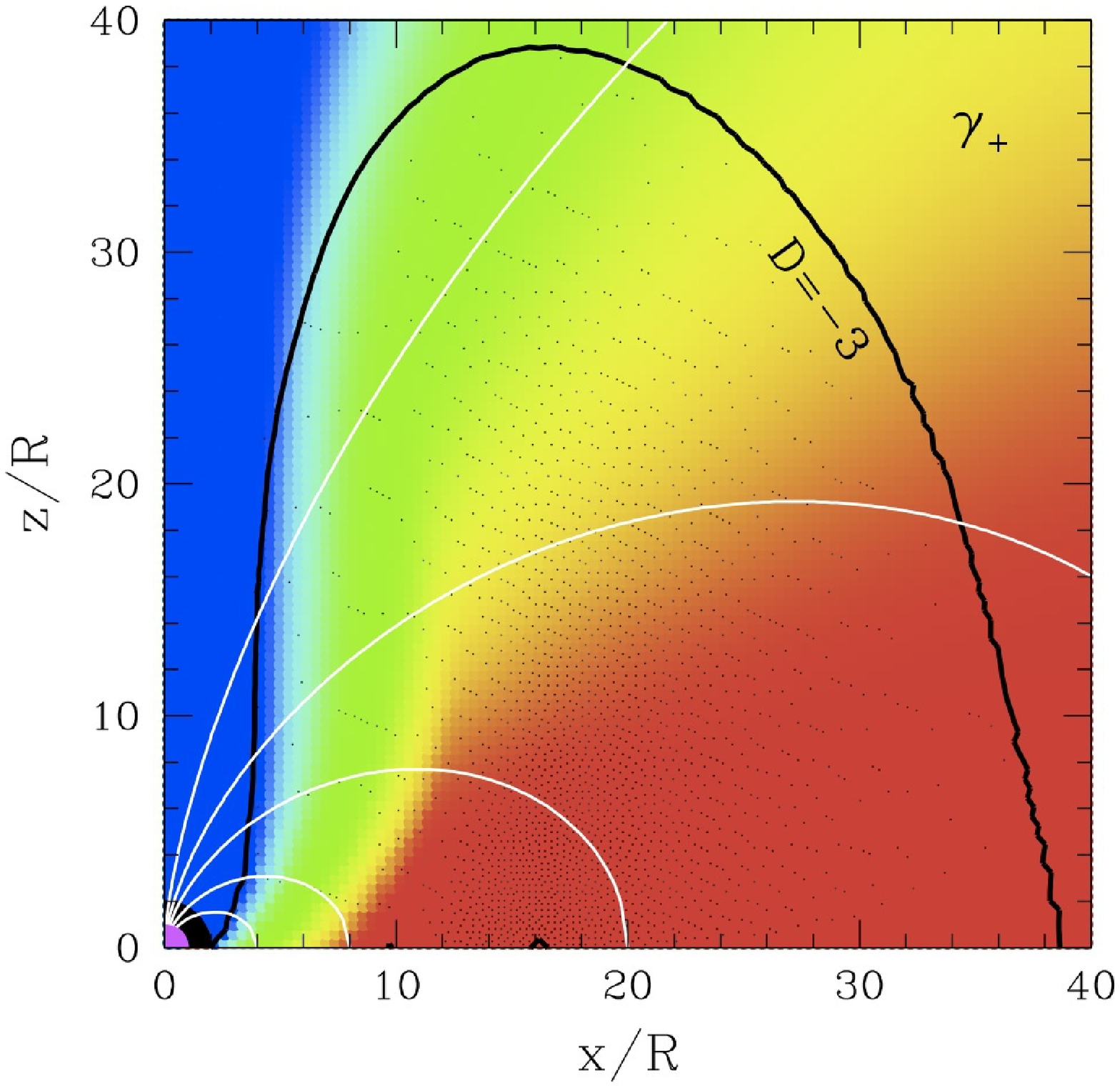} 
\hspace*{1cm} & 
\includegraphics[trim = 2cm 0cm 1cm 0cm, width=0.42\textwidth]{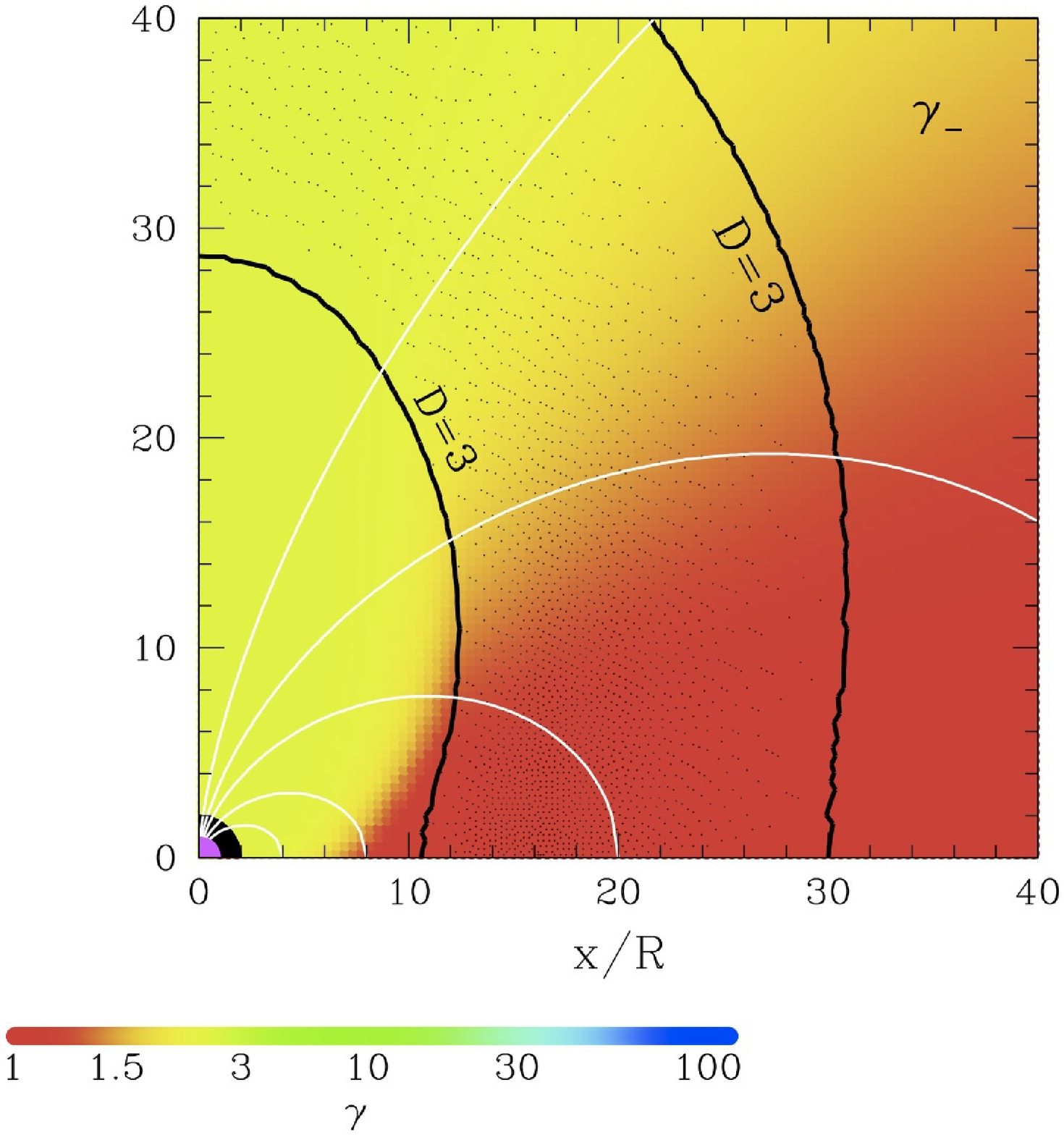}
\end{tabular}
\caption{
Lorentz factors $\gamma_+$ (left panel) and $\gamma_-$ (right panel)
in the two-fluid model of the $e^\pm$ flow. In this example, the electric 
current is carried by the $e^\pm$ outflow of a fixed multiplicity $\M=50$.
The plasma is injected at radius $r=2R$ and outflows along the magnetic 
field lines (white curves). The flow is illuminated by the star with temperature 
$\kB T=0.5$~keV (magenta circle at the origin), and the radiation exerts the 
forces $\F(\gamma_\pm)$ on the positron ($+$) and electron ($-$) fluids.
The Lorentz factors $\gamma_+$ and $\gamma_-$ change as the 
flow enters the drag-dominated region $|\D|\gg 1$. The region $|\D|>3$ 
is shown by the thick black curve and shadowed in black.
$\D<0$ for positrons ($\gamma_+>\gamma_\star$) and $\D>0$ for electrons
($\gamma_-<\gamma_\star$).
~The radiative drag stops the plasma in the equatorial plane
outside $\sim 8R$. A nearly dipole magnetic field (weakly 
twisted) with $B_{\rm pole}=10^{15}$~G is assumed in this example.
$R$ is the neutron-star radius.
} 
\end{figure*}

\subsection{Sample numerical model}

Suppose that $e^\pm$ plasma is injected near the star with a given
multiplicity, e.g. $\M=50$, and a given high Lorentz factor, e.g. 
$\gamma_+=100$. The corresponding $\gamma_-$ is determined by 
\Eq~(\ref{eq:j}).
Suppose that the plasma is illuminated by the blackbody radiation of 
the star of temperature $\kB T=0.5$~keV and neglect radiation from the 
magnetosphere itself. This approximation  
is valid only for optically thin magnetospheres, which are considered here 
for simplicity.
A more detailed model will be developed in \Sects~5 and 6, which will take 
into account radiation scattered in the magnetosphere;
then radiation field is not central, and we will have to solve 
radiative transfer to determine the flow momentum. 

An explicit expression for $\F(\gamma)$ exerted by the central radiation is 
given in Appendix~B (\Eq~\ref{eq:f}).
The steady-state solution for $\gamma_\pm$ in the outer corona  
can be found by integrating \Eq~(\ref{eq:gam_2F}) along the magnetic 
field lines. The result is shown in Figure~3. The relativistic outflow is injected 
near the star and initially weakly interacts with the radiation; then it enters the 
drag-dominated region $|\D|\gg 1$. The solution is not 
sensitive to the precise radius of $e^\pm$ injection as long as it is 
small enough, before the plasma enters the drag-dominated region.

The electric field in the region of $|\D|\gg 1$ is given by 
$e\Epar\approx -\F(\gamma_+)$, and the 
corresponding longitudinal voltage established in the outer corona
is found by integrating $\F(\gamma_+)$ along the field line,
$e\Phi_e\approx -\int \F(\gamma_+) \,dl$.
Its typical value for the model in Figure~3 is $\sim 10^7$~V.
Flows with lower $\M$ develop stronger electric fields, however in all 
cases of interest ($\M\gg 1$) the drag-induced voltage is below $10^9$~V.

The calculations shown in Figure~3 assume that the magnetospheric plasma 
is everywhere optically thin. This is not so for real magnetars. 
Thompson et al. (2002) showed that the characteristic optical depth $\tau$ of 
a strongly twisted magnetosphere (twist amplitude $\psi\sim 1$) is comparable 
to unity. When the large pair multiplicity $\M$ is taken 
into account, the estimate changes to
\beq
\label{eq:tau}
   \tau\sim \frac{\M\psi}{\beta_\pm} \gg 1.
\eeq
This estimate describes the optical depth seen by photons that can be 
resonantly scattered by the flow, i.e. the resonance condition 
$\gamma(1-\beta\cos\ang)\omega=\omega_B$ is satisfied somewhere 
along the photon trajectory. The large $\tau$ implies the presence of  
scattered radiation in the magnetosphere, which is quasi-isotropic
rather than central.
This increases the drag exerted on the outflow and reduces $p_\star$.
One may also expect a self-shielding effect: 
the drag force $\F$ experienced by an electron (or positron) is reduced 
by the factor of $\tau^{-1}$. 
The problem of self-consistent outflow dynamics will be solved in \Sects~5 and 6.  
The main feature seen in Figure~3 will persist in the full self-consistent solution: 
the outflow is strongly decelerated (and drag-dominated) in the equatorial region 
at $r\sim (10-20)R$.


\section{Two-stream instability, anomalous resistivity, and radio emission}
\label{sec:inst}

\subsection{Two-stream instability}

The two-fluid flow with $v_+>v_-$ 
is prone to the two-stream instability (e.g. Krall \& Trivelpiece 1973). 
The growth rate of the instability is obtained from the dispersion relation for 
Langmuir modes with frequency $\omega$ and wavevector $k$, which can 
be derived by considering perturbations $\delta\gamma_\pm$ 
of the two-fluid system and using continuity, Euler, and Poisson equations, 
\begin{equation}
\label{eq:disp}
    1-\frac{\omega_+^2}{\gamma_+^3(\omega - kv_+)^2} 
     - \frac{\omega_-^2}{\gamma_-^3(\omega - kv_-)^2} = 0,
\end{equation}
where $\omega_\pm^2=4\pi n_\pm e^2/m_e$. The plasma is nearly neutral,
$n_+=n_-$, and hence $\omega_+=\omega_-$. 
\Eq~(\ref{eq:disp}) has a solution $\omega(k)$ with a positive imaginary part that 
describes an unstable mode. The solution simplifies in the following two limits:

\noindent
(1) $\gamma_+-\gamma_-\ll \gamma_\pm$, which is valid when 
$\M\gg \gamma_\pm^2$ (see \Eq~\ref{eq:j}).
Then the flow is convenient to view in its center-of-momentum frame that moves 
with $\gamma\approx(\gamma_++\gamma_-)/2$. In this frame, the two 
fluids with densities  $\tilde{n}_\pm=\gamma^{-1} n_\pm$ move in the opposite 
directions with non-relativistic velocities $\tilde{v}_\pm=\pm\tilde{v}$, so the 
dispersion relation~(\ref{eq:disp}) simplifies.  It gives the most unstable mode 
$\tilde{k}=(\sqrt{3}/2)\,\tilde{\omega}_\pm/\tilde{v}$ with a growth 
rate $\tilde{\Gamma}=\tilde{\omega}_\pm/2$. 

\noindent
(2)  $\gamma_+/\gamma_-\gg 1$. Then the contribution of positrons to 
the dispersion relation is small compared to that of the electrons. The 
growth rate of the instability is given by 
$\Gamma\approx \gamma_-^{-1/2}\gamma_+^{-1}\omega_p$
(e.g. Lyubarsky \& Petrova 2000).

This estimate gives the characteristic length-scale of the instability $c/\Gamma$,
which is much shorter than the electron free path to resonant scattering, 
$\lambda_{\rm sc}$. Hence the radiation drag and the 
induced electric field $e\Epar=\pm\F(p_\pm)$ are unable to lock the positive
and negative charges at the momenta $p_+$ and $p_-$ 
calculated in the two-fluid model. The instability will generate plasma 
oscillations that should broaden the momentum distribution so that particles fill the 
region $p_-<p<p_+$. 

The generated plasma oscillations may be expected to introduce
an anomalous resistitivity. The fluctuating $\Epar$ in the oscillations
creates a stochastic force that tends to reduce the free-path of a
charged particle.
A simplest estimate suggests that this effect could be very strong.
Suppose a substantial fraction of the energy density of the flow
$\gamma m_e c^2 n$ is given to plasma oscillations.
Then the characteristic electric
field is $\Epar\sim (8\pi \gamma m_e c^2 n)^{1/2}$; it is much 
stronger than $\Epar$ in the radiatively locked two-fluid model,
however it is irregular and quickly changes sign. Suppose that the
stochastic electric force exerted on the particle randomly changes sign on
a timescale $\Delta t\sim\omega_p^{-1}$, where $\omega_p$ is the plasma
frequency. The stochastic $\Epar$ gives the momentum kicks
$\Delta P\sim e\Epar \omega_p^{-1}$ and causes diffusion of particles in
the momentum space with the diffusion coefficient
\beq
\label{eq:diff}
  D_p\sim \frac{(\Delta P)^2}{\Delta t}\sim \frac{(e\Epar)^2}{\omega_p}.
\eeq
Diffusion in momentum space $p^2(t)\sim D_p t$ implies  
a small free-path of the particle,
$\lambda\sim (\gamma\beta)^2 m_e^2c^3\omega_p/(e\Epar)^2$,
much smaller than the mean free path to resonant scattering.
Thus, a large anomalous resistivity could, in principle,
be possible, and then a large longitudinal voltage would be generated
to maintain the electric current.

\subsection{Numerical experiment}

To explore the role of the two-stream instability and anomalous
resistivity, we designed the following numerical experiment. 
Keeping in mind that particles around magnetars can flow only along the 
magnetic field lines, consider the simple one-dimensional problem. 
Suppose an $e^\pm$ beam is continually injected at the boundary
$z=0$ of the computational box $0<z<L$. The rate of
electron injection is smaller than the rate of positron injection,
so that the flow carries current $j>0$. Positrons are injected with fixed
$p_+$ and electrons with fixed $p_-<p_+$;
we chose $p_+=7$ and $p_-=0.5$. The simulation
keeps track of the flow in the computational box of length
$L\sim 150c/\omega_p$. 
The escape boundary condition is implemented at $z=L$.
The flow is simulated as a large collection of individual particles ($N\sim 10^6$)
that move in their collective electric field (see Beloborodov \& Thompson [2007] 
for details of the numerical method). The simulation was run for a time of 
$t\sim 100L/c$, long enough to see the quasi-steady behavior.
                                                                                
The results of the simulation are as follows. As expected, strong plasma 
oscillations develop and persist in the flow, as the instability is continually 
fed by the injection of the two streams at $z=0$. 
The fluctuating electric field reaches very high amplitudes that could 
reverse a particle with Lorentz factor $\gamma_+$ on a short scale,
much shorter than $L$. A snapshot of $\Epar(z)$ is shown in Figure~4. 
The integral of the electric field over $z$ determines the voltage $\Phi_e$ 
between the two boundaries of the computational box. The measured 
$\Phi_e$ in the simulation fluctuates in time  (Figure~5), and we calculated 
its value averaged over $100L/c$. This value turns out to be very small, 
$e\Phi_e\sim 0.1\gamma_+m_ec^2$. Thus, the measured anomalous 
resistivity is small, in sharp contrast with the simplest estimate~(\ref{eq:diff}) 
that would predict $\lambda\ll L$ and hence $e\Phi_e\gg \gamma_+m_ec^2$.

\begin{figure}
\vspace*{-2.2cm}
\epsscale{1.1}
\plotone{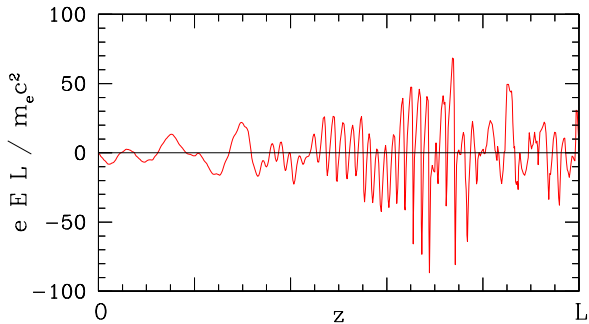}
\vspace*{-0.5cm}
\caption{Snapshot of electric field in the simulation box.}
\end{figure}
\begin{figure}
\epsscale{1.10}
\plotone{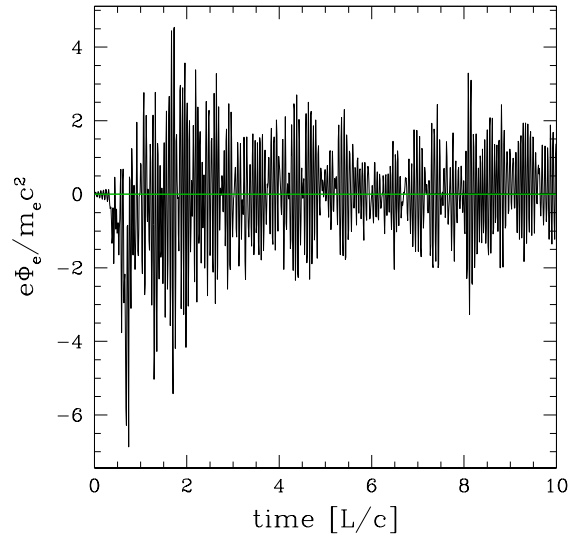}
\caption{Voltage across the simulation box as a function of time.}
\end{figure}
\begin{figure}
\epsscale{1.00}
\plotone{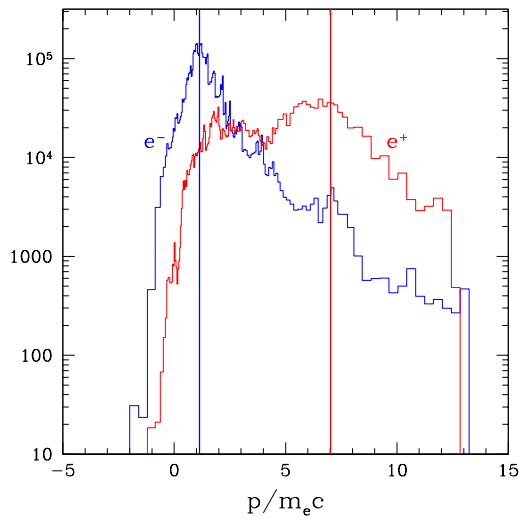}
\caption{Snapshot of the momentum distributions of $e^+$ (red) and $e^-$ (blue).
Vertical lines show the injection $p_\pm$ at $z=0$.}
\end{figure}
                 
The failure of the estimate~(\ref{eq:diff}) is related to the assumption
that the stochastic electric force applied to the particle is random,
uncorrelated on timescales longer than $\omega_p^{-1}$.
The numerical simulation indicates that this assumption is incorrect.
Apparently, a complicated time-dependent pattern is
organized in the phase space, which allows the charges to
find small-resistance paths through the waves of $\Epar$
and conduct the current at a low net voltage and a low dissipation rate.

One limitation of the numerical experiment should be noted:
the one-dimensional computational box allows no dependence of $\Epar$ 
on the transverse coordinates $x$ and $y$, which excludes the coupling of 
Langmuir waves to the transverse electro-magnetic modes. A two-dimensional
(or a complete three-dimensional) simulation will be needed to explore the 
role of the transverse modes. We anticipate that a low anomalous resistivity
will be found in the full simulations.  A high resistivity would imply a quick 
dissipation of magnetospheric currents, which would produce a high luminosity 
and quickly erase the magnetic twist. This is not supported by observations of 
magnetars. The observed luminosities and evolution timescales are consistent 
with the model neglecting the anomalous resistivity, where voltage is controlled 
by the threshold of the $e^\pm$ discharge, $e\Phi_e\sim 10^9-10^{10}$~V.

The numerical simulation shows that the two-stream instability significantly 
changes the momentum distributions of electrons and positrons from the 
injected delta-functions $\delta(p-p_-)$ and $\delta(p-p_+)$. The $e^\pm$ 
distributions are broadened so that they fill the region between the injection 
momenta $p_+$ and $p_-$ (Figure~6).

\subsection{Low-frequency emission}

An important implication of the two-stream instability is the excitation of a 
strong plasma turbulence that can generate coherent low-frequency radiation. 
The two-stream instability is often considered in pulsar models as a mechanism 
feeding radio emission from the open field-line bundle 
(e.g. Sturrock 1971; Cheng \& Ruderman 1977).
A related model invoking radiative drag was considered 
by Lyubarsky \& Petrova (2000).\footnote{They suggested that a broad   
      momentum distribution of relativistic particles in the open 
      field-line bundle  can evolve into a two-hump  
      distribution as a result of resonant scattering losses, as the lower energy 
      particles lose their energy faster than the more energetic ones. In contrast,
      the two-fluid flow described in \Sect~3 is shaped by the induced $\Epar$ so that 
      it sustains the electric current $j$; without $\Epar$ radiative drag 
      would equalize the velocities of all particles at $v_\star$.
      }
In the $e^\pm$ flows around magnetars, the two stream instability is naturally driven 
by the strong electric current in the system that tends to lock itself in the two-fluid
configuration as described in \Sect~3. The instability is continually pumped by 
the radiative drag in the dense radiation field of the magnetar.
The generated low-frequency radiation has the best chance to escape near the
magnetic axis, where the plasma density is lowest and its Lorentz factor is highest.
Note that the two-stream instability operates on the closed (twisted) field lines,
which carry significant magnetic flux. This allows the low-frequency emission to be 
unusually bright, even though the voltage $\Phi_e\sim 10^9-10^{10}$~V is small by 
the ordinary-pulsar standards.

Radio pulsations have been detected and studied in detail in two magnetars
XTE~1810-197 and 1E~1547.0-5408 (Camilo et al. 2006; 2007). 
The estimated radio luminosity $L_r\sim 10^{30}$~erg~s$^{-1}$ 
requires a sufficiently high ohmic dissipation rate, $I\Phi_e>L_r$, which 
cannot be generated in the open field-line bundle unless $\Phi_e$ exceeds 
$L_r/\Iop\sim 10^{11}\varepsilon_r^{-1} L_{r,30}$~V. Here $\varepsilon_r$ is the 
efficiency of radio emission, $\Iop$ is the electric current circulating in the 
open bundle, $\Iop\sim c\mum/\RLC^2$, and $\RLC=cP/2\pi\sim 10^{10}$~cm 
is the light cylinder for a magnetar rotating with the period $P$ of a few seconds. 
As we argued above, voltage is likely near the threshold of $e^\pm$ discharge, 
which gives $\Phi_e\ll 10^{11}$~V. 

Therefore, we conclude that the electric current associated with observed radio 
emission is large, $I\gg \Iop$, and should flow in the closed (twisted) 
magnetosphere, giving a bright and relatively broad radio beam. This is 
consistent with the unusually broad radio pulses of magnetars, much broader 
than the typical pulse of ordinary pulsars with similar periods 
(Camilo et al. 2006; 2007).
Note also that the plasma density in the twisted closed magnetosphere 
reaches much higher values than in the open field-line bundle, 
and the plasma frequency may approach the infrared band.
This may explain the observed hard radio spectra.

One could consider the possibility that the radio luminosity of magnetars is 
generated by enhanced dissipation in the open field-line bundle and its immediate
vicinity. Thompson (2008b) suggested that the diffusion of magnetic 
twist $\psi$ in the closed magnetosphere initiates a strong Alfv\'enic turbulence 
near the light cylinder, with a high dissipation rate. 
This picture assumes that the magnetic twist tends to spread due to ohmic 
diffusion, as observed in normal laboratory plasma with a finite resistivity. 
However, later work (Beloborodov 2009; 2011a) showed that the twists in 
neutron-star magnetospheres evolve differently: the twist is erased 
``inside out'' rather than spreads diffusively. The twist evolution 
$\partial\psi/\partial t$ is controlled by 
voltage induced along the magnetic field lines, $\Phi_e$, so that 
$\partial\psi/\partial t\propto d\Phi_e/d\Fm$, where $\Fm$ is the magnetic flux 
function labeling the field lines ($\Fm=0$ on the magnetic dipole axis).
An increased voltage near the axis would imply a large negative 
$d\Phi_e/d\Fm$, which implies a large negative $\partial\psi/\partial t$, 
i.e. rapid untwisting. Thus, the high-$\Phi_e$ twist near the open field-line 
bundle cannot be sustained.
The ohmic effects in the closed magnetosphere can pump the twist near the 
open bundle only if $\Phi_e$ is {\it reduced} toward the magnetic dipole axis, 
i.e. $d\Phi_e/d\Fm>0$. Then the twist pumping continues until the expanding 
cavity $j=0$ reaches the open bundle;
a snapshot of this evolution is shown in Figure~1. The pumping of $\psi$ 
leads to outbursts and spindown anomalies (Parfrey et al. 2012, 2013). 

Our scenario for the low-frequency emission from magnetars may be 
summarized as follows. The emission is generated by the two-stream instability 
on the twisted closed field lines with the apex radii $\Rmax$ such that 
$R\ll\Rmax\ll\RLC$. These field lines carry a large electric current 
$I\sim \psi c\mum/\Rmax^2\gg \Iop$ and a modest voltage $\Phi_e\sim 10^9-10^{10}$~V.
The high plasma density and the broad beam of radiation expected on these
field lines explain the unusual radio pulsations of magnetars.


\section{Dynamics of outflow with a broad momentum distribution}

\subsection{Waterbag model}

The plasma instability discussed in \Sect~4 is generated by the gradient of 
the distribution function $d\f/dp$, and the feedback of the excited plasma 
waves tends to make the distribution flatter.
The numerical simulation in \Sect~\ref{sec:inst}
illustrates how two interpenetrating cold fluids of $e^+$ and $e^-$ with 
$\f(p)=(1/2)[\delta(p-p_-)+\delta(p-p_+)]$ quickly evolve into a state 
with a broad and smooth $\f(p)$. Below we design a simple modification of 
the two-fluid model that takes this effect into account. 

The simplest model has a top-hat distribution function. 
In plasma physics, this approximation is often called ``waterbag'' model.
Like the two-fluid model, the outflow is described by two parameters 
$p_\pm$ (or $\beta_\pm$). However, now instead of two 
delta-functions we require the $e^\pm$ distribution to be flat 
between $p_-$ and $p_+$,
\beq
  \f(p)=\left\{\begin{array}{cc}
     (p_+-p_-)^{-1} &  p_-<p<p_+ \\
           0        &  p<p_- \mbox{~~or~~} p>p_+ 
               \end{array}
        \right.
\eeq
This distribution includes both electrons and positrons.
The plasma must be nearly neutral, $n_+=n_-$, and the total density of 
particles $n=n_++n_-$ is given by
\beq
   n=\frac{\dN}{\bar{\beta}c}=\frac{\M j}{e\,\bav c},
\eeq
where $\dN=\M j/e$ is the particle number flux and $\bav$ is 
the average velocity, 
\beq
\label{eq:bav}
   \bar{\beta}=\int \beta(p)\,\f(p)\,dp
              =\frac{\gamma_+-\gamma_-}{p_+-p_-}.
\eeq

\begin{figure}
\epsscale{1.1}
\plotone{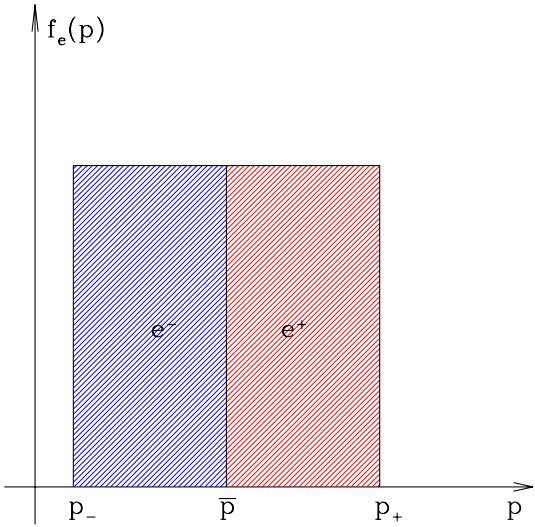}
\caption{Waterbag distribution of minimum width.}
\end{figure}

Similar to the two-fluid model, $p_+$ and $p_-$ are not independent,
as the outflow must carry current $j$ with a given multiplicity $\M$; this would 
be impossible if, for instance, $p_+=p_-$. The minimum possible width of the 
distribution function (i.e. minimum $p_+-p_-$) is achieved if all negative 
charges are slower than all positive discharges, as shown in Figure~7. 
We will adopt this idealized momentum distribution in our numerical simulations.
This is a rather crude approximation to more realistic distributions of $e^+$ 
and $e^-$, which overlap in momentum space (cf.~Figure~6).\footnote{
      The minimum-width waterbag model may be particularly crude near the 
      star where the outflow just begins to experience significant radiative drag.
      The faster positive charges will experience drag first while the slower
      negative charges still move freely. As $p_+$ decreases, the minimum-width
      model would require an increase in $p_-$. In reality, $p_-$ may not 
      react to the reduced $p_+$ until particles with $p=p_-$ also begin to 
      experience the drag force (which is positive as $p_-<p_\star$). After this point, 
      the outflow may not be far from the ``minimum-width'' waterbag state.}
Nevertheless, it is useful as it allows 
one to explore all basic features of the $e^\pm$ outflow using a concrete 
relation between $p_+$ and $p_-$.
This relation is determined by the outflow multiplicity $\M$.
It is easy to show that the generalization of the two-fluid \Eq~(\ref{eq:j}) to 
$e^\pm$ flows with any distribution function is given by
\beq
  1-\frac{\bav_-}{\bav_+}=\frac{2}{\M+1},
\eeq
where $\bav_+$ and $\bav_-$ are the average velocities of the positive and 
negative charges, respectively. 
For the waterbag distribution shown in Figure~7 this condition gives
\beq
\label{eq:j1}
    \frac{\bav_-}{\bav_+}
        =\frac{\gamma(\pav)-\gamma_-}{\gamma_+-\gamma(\pav)}
        =1-\frac{2}{\M+1}, \qquad \pav=\frac{p_-+p_+}{2},
\eeq
where $\gamma(\pav)=(1+\pav^2)^{1/2}$.
\Eq~(\ref{eq:j1}) determines the relation between $p_+$ and $p_-$ 
(for a given $\M$); it is shown in Figure~8.

\begin{figure}
\epsscale{1.1}
\plotone{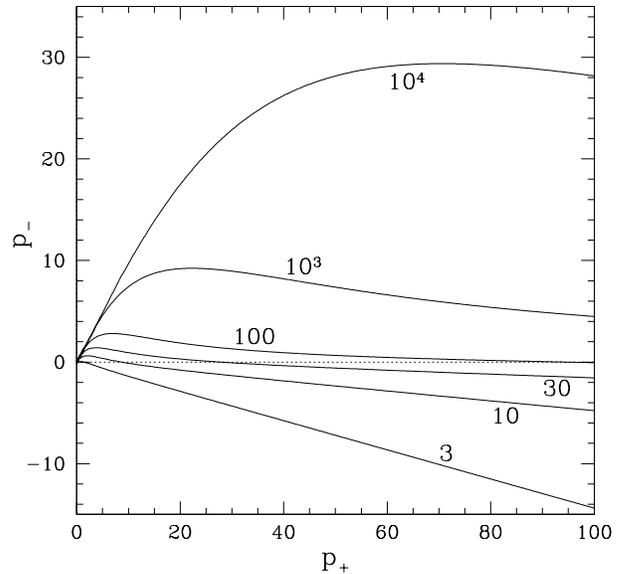}
\caption{Relation between $p_+$ and $p_-$ for the waterbag model shown in 
Figure~7. Each curve describes an outflow of a given multiplicity $\M$, which is 
indicated next to the curve; $p_-=0$ is indicated by the dotted line.
} 
\end{figure}

Taking into account the relation between $p_+$ and $p_-$, the outflow 
has essentially one degree of freedom besides the multiplicity $\M$.
One can chose, e.g., $p_+$ as an independent variable, or any convenient 
combination of $p_+$ and $p_-$, $\zeta(p_+,p_-)$, that is independent from 
$\M(p_+,p_-)$. Below we will choose variable $\zeta$ and formulate the 
dynamic equation for $\zeta$; it will describe how the interaction with radiation 
governs the outflow dynamics along the magnetic field lines.

\subsection{Momentum equation}

Since the outflow is bound to move along the magnetic field lines, we will 
need the projection of momentum equation onto $\bB$. It is convenient 
to write this equation in a covariant form that is valid in any curved coordinate 
system $x^i$ (e.g. spherical coordinates). In a steady state, the momentum 
equation reads 
\beq
\label{eq:mom}
    \frac{B_i}{B}\,\nabla_k T^{ik}=\frac{dP}{dt\,dV},
\eeq
where $dP/dtdV$ is the rate of momentum exchange with radiation 
(per unit volume), $\nabla_k$ is the covariant derivative (indices $i$, $k$ 
run from 1 to 3), and $T^{ik}$ are the components of the plasma stress tensor,
\beq
\label{eq:Tik}
   T^{ik}=m_ec^2\,\int u^i u^k\,\frac{dn}{\gamma}
           =n\,m_ec^2\,\frac{B^iB^k}{B^2}\int \frac{p^2}{\gamma}\,\f(p)\,dp.
\eeq
Here $u^i=p\,B^i/B$ are the spatial components of the four-velocity vector of
a particle with dimensionless momentum $p$, $dn=n \f(p) dp$ is the number 
density of particles with momenta $(p,p+dp)$, and $dn/\gamma$ is the 
corresponding density in the frame moving with $\gamma(p)=(1+p^2)^{1/2}$. 
Then the left-hand side of \Eq~(\ref{eq:mom}) takes the form,
\begin{eqnarray}
\nonumber
   \frac{B_i}{B}\,\nabla_k T^{ik} & = & \frac{B_i}{B}\,\nabla_k
                 \left(n\,m_ec^2\,\frac{B^iB^k}{B^2}\,\overline{p\beta}\right) \\
        & = & m_ec^2\,B^k\nabla_k\left(\frac{n}{B}\,\overline{p\beta}\right),
\label{eq:lhs}
\end{eqnarray}
where we used $B_i\nabla_k(B^iB^k/B)=0$ (which follows from $\nabla_k B^k=0$).
The directional derivative $B^k\nabla_k$ equals $B\,d/dl$ where $l$ is length 
measured along the magnetic field line. \Eq~(\ref{eq:lhs}) can be further simplified
using $n\bav/B\M=\const$, which follows from the relation
\beq
    n=\M\,\frac{j}{ec\bav},
\eeq
and $j/B=const$ (cf.~\Eq~\ref{eq:jpm}). This gives
\beq
\label{eq:lhs1}
   \frac{B_i}{B}\,\nabla_k T^{ik}=m_ec^2\,n\,\frac{\bav}{\M}\,\frac{d}{dl}
                \left(\frac{\M}{\bav}\,\overline{p\beta}\right).
\eeq
When $\M=\const$ (i.e. no new pairs are created), $\M$ cancels out.
Finally, the substitution of \Eq~(\ref{eq:lhs1}) to \Eq~(\ref{eq:mom})  gives
the momentum equation in the following form,
\beq
\label{eq:dyn1}
    \frac{d\zeta}{dl}=\frac{\FF}{\bav\,m_ec^2},  \qquad 
                \zeta\equiv\frac{\overline{p\beta}}{\bav},
\eeq
where $\FF=n^{-1} dP/dtdV$ is the average force exerted by radiation 
{\it per particle}. If $\FF=0$ then $\zeta$ remains 
constant along the field lines, which implies $p_\pm=\const$ --- the 
momentum distribution remains unchanged along the flow.

For the waterbag model, $\bav$ is given by \Eq~(\ref{eq:bav}).
The quantity $\overline{p\beta}$ can also be expressed in terms 
of $p_\pm$, using the indefinite integral
\beq
   \int p\beta\,dp=\int p\,d\gamma=\frac{p\gamma}{2}
           -\frac{1}{4}\ln\left(\frac{1+\beta}{1-\beta}\right)+\const.
\eeq
This gives
\beq
\label{eq:zeta}
   \zeta=\frac{1}{\gamma_+-\gamma_-}
   \left[\frac{1}{2}(p_+\gamma_+ - p_-\gamma_-)
           -\frac{1}{4}\ln \frac{(1+\beta_+)(1-\beta_-)}{(1-\beta_+)(1+\beta_-)}\right].
\eeq 
In our numerical simulations, we use $\zeta$ 
as the variable that describes the dynamical state of the outflow, 
and solve the differential \Eq~(\ref{eq:dyn1}) for $\zeta$. The values of $p_\pm$ 
that correspond to a given $\zeta$ and $\M$ are found from \Eqs~(\ref{eq:j1})
and (\ref{eq:zeta}).

The force $\FF$ experienced by an outflow with a given $\zeta$ is determined 
by the local radiation field, which is described by the intensities in the two 
polarization modes $I_\perp(\omega,\nn)$  and $I_\parallel(\omega,\nn)$
(Appendix~A).
In general, the force acting on a plasma with a distribution 
function $\f(p)$ is given by \Eq~(\ref{eq:dP3}). 
Force $\FF$ is easily calculated in the simple case of an optically thin 
magnetosphere exposed to the central blackbody radiation (Appendix~B). Then
the right-hand side of \Eq~(\ref{eq:dyn1}) is a known function of $\zeta$ and
it is straightforward to numerically solve this equation; the results are similar to
the two-fluid model described in \Sect~4.
The optically thin approximation is, however, invalid for active magnetars. 
The radiation is not central; instead $I_\perp$ and $I_\parallel$ must be 
calculated self-consistently, together with the outflow dynamics. This requires 
radiative transfer simulations.

Note that we assume here that $p_\pm>0$, so that all particles move
away from the star until they reach the equatorial plane and 
disappear (annihilate) there. This approximation is reasonable for the flow
along the extended field lines with apex radii $\Rmax\simgt 10R$.
Near the apexes, i.e. near the equatorial plane $\theta=\pi/2$,
the radiative drag enforces $p_\pm\approx p_\star\ll 1$, 
creating a dense layer of slow particles where they annihilate (Fig.~2). 
A more general model of the flow could allow the particles to cross the 
equatorial plane and enter the opposite hemisphere.
By symmetry, this would be equivalent to the reflection 
boundary condition, i.e. the mirror image of the outflow approaching the 
equatorial plane would emerge from the equatorial plane. Then the 
distribution function $\f(p)$ must extend to negative $p$.
This modification would be required
for the plasma flow on field lines with small $\Rmax$, 
where radiative drag is less efficient and the plasma can cross the equatorial 
plane with a large $p$. In this paper, we focus on the field lines with 
$\Rmax\simgt 10$, where this does not happen. 
The field lines with small $\Rmax$ are assumed to form a cavity with 
$j=0$ and a negligible plasma density (Figure~1).


\section{Self-consistent radiative transfer}

The problem of radiative transfer in a relativistically moving $e^\pm$ plasma 
whose velocity is controlled by the radiation field 
is not unique to magnetars. A similar situation may occur in accretion disk 
outflows (Beloborodov 1998) and gamma-ray bursts (Beloborodov 2011b). 
The strong radiative drag (measured by the coefficient $\D\gg 1$,  
\Eq~[\ref{eq:drag}])
was previously shown to simplify the problem, as it forces the plasma to 
keep the 
saturation momentum $p_\star$ such that the net radiation flux vanishes in 
the plasma rest frame. 
This ``equilibrium'' transfer has one additional integral compared with the 
classical Chandrasekhar-Sobolev transfer problem for a medium at rest.
The transfer in magnetar magnetospheres has, however, two special features
that complicate the problem.
First, the electric current and plasma instabilities imply additional (electric) 
forces that broaden the momentum distribution around $p_\star$, as 
discussed in \Sects~4 and 5. Therefore, the equilibrium condition $p=p_\star$
is not satisfied even where $\D\gg 1$. Second, opacity 
is dominated by {\it resonant} scattering, whose rate is sensitive to the 
particle momentum. 
Below we develop a method to solve the self-consistent transfer for magnetars.

\subsection{Monte-Carlo technique and the ``virtual beam'' method}

Radiative transfer in a magnetosphere filled with plasma with given 
parameters can be calculated using the standard Monte-Carlo technique 
(e.g. Fern\'andez \& Thompson 2007; Nobili, Turolla \& Zane 2008).
Blackbody photons are injected into the magnetosphere at the neutron-star
surface and their trajectories are followed until they escape the magnetosphere.
We implement this method using the scattering opacity given in Appendix~A
and keeping track of the photon polarization, which can switch in the scattering 
events. 

Calculation of transfer in an outflow with self-consistent dynamics is a more 
ambitious goal, as the plasma parameters are not known in advance. 
A natural approach is iterative. One can start with a trial outflow,
calculate radiative transfer to find the radiation intensity that would correspond
to this outflow, and then re-calculate the outflow dynamics in the obtained 
radiation field. These iterations can be repeated until they converge. 
The problem is simplified for the waterbag plasma model 
(\Sect~5) as the outflow is described by one dynamic variable $\zeta$. 
As the first trial initiating the iterations one can take the outflow solution 
$\zeta(r,\theta)$ for the optically thin magnetosphere exposed to the central 
blackbody radiation.

One then encounters the following difficulty. 
For the calculation of next iteration of outflow dynamics, one needs to know 
$I_\perp(\br,\omega,\nn)$ and $I_\parallel(\br,\omega,\nn)$ everywhere in the 
magnetosphere. In axisymmetric magnetospheres, the radiation intensity
is a function of five variables: two for location (radius $r$ and polar 
angle $\theta$), one for spectrum ($\omega$), and two for angular distribution 
(unit vector $\nn$ is described by two angles). 
To determine intensities $I_\perp$ and $I_\parallel$ with sufficient accuracy,
one has to introduce a five-dimensional grid and 
accumulate large photon statistics for each grid cell during the Monte-Carlo 
simulation of radiative transfer. A grid of size $N$ for
 each of the five variables has $N^5$ cells. Accumulation of 
photon statistics in each cell requires the calculation of a huge number of 
Monte-Carlo realizations of the photon trajectory in the magnetosphere.
This is expensive and gives poor accuracy.

There is, however, a more efficient method that does not require the 
knowledge of intensities $I_{\perp,\parallel}(\br,\omega,\nn)$. They are not, 
in fact, needed for the calculation of outflow dynamics. What enters the 
momentum \Eq~(\ref{eq:dyn1}) is the average force $\FF(r,\theta)$, which 
may be tabulated in the two-dimensional space of $r,\theta$. This force can 
be calculated directly during the Monte-Carlo simulation of radiative transfer
if we find a way to evaluate the contribution of each simulated photon to 
$\FF(r,\theta)$ as it propagates through the magnetosphere.

This can be achieved if we imagine that the photon is replaced by a beam 
of radiation. As we follow the photon along its 
trajectory, we can calculate the force that would be applied by the imaginary 
beam to any given outflow that can be imagined in the magnetosphere. In the
waterbag model, the outflow is described by one dynamical parameter $\zeta$, 
and so the force applied by the beam can be tabulated on a grid of $\zeta$.

Once we know how to calculate the force created by each photon trajectory 
in our Monte-Carlo simulation, we should be able to average it over all 
simulated photons and thus accurately evaluate $\FF(r,\theta,\zeta)$. 
This gives the force that {\it would be} applied by our radiation 
field to an outflow with any given $\zeta$ at any point $r,\theta$.
At the next iteration step $\FF(r,\theta,\zeta)$ is used to 
   obtain the new outflow solution $\zeta(r,\theta)$
by integrating the momentum equation $d\zeta/dl=W(l,\zeta)$, 
where $W=\FF/\bav m_ec^2$ (\Eq~\ref{eq:dyn1}). 
Then the Monte-Carlo simulation can be repeated to calculate radiative transfer 
in the new outflow and find  $\FF(r,\theta,\zeta)$ for the new radiation field.
These steps can be repeated until the outflow solution $\zeta(r,\theta)$ 
converges, i.e. remains practically unchanged by new iterations.

The concrete implementation of this strategy is as follows. Let $\Lth$ be the 
thermal luminosity of the star and $\dN=\Lth/2.7kT$ be the number of 
photons emitted by the star per unit time. In our 
Monte-Carlo simulation we follow $\KMC\sim 10^7$ random photon trajectories. 
This can be thought of as dividing $\dN$ into $\KMC$ 
random monochromatic beams. Each beam has a random start at the star 
surface (the photon energy is drawn from the Planck distribution) and follows 
one random realization of the photon trajectory, which can involve multiple 
scattering events in the magnetosphere. The photon number flux in each 
monochromatic beam is $\dN_b=\dN/\KMC$, and the energy flux in the beam is 
\beq
    \dE_b=\dN_b\, \hbar\omega.
\eeq
Note that $\dN_b=\const$ along the beam (i.e. along the photon trajectory 
in the Monte-Carlo simulation) while the photon energy $\hbar\omega$ 
changes after each scattering in the magnetosphere.
The collection of $\KMC$ beams represent the state of the radiation field
around the star.
 
As we follow each realization of the photon trajectory, in parallel we calculate 
the force applied by the virtual beam to the plasma.
We can imagine that the beam has a small cross section $A$ (it will cancel 
in the final result) and flux density $F=\dE_b/A$.
\Eq~(\ref{eq:dP2}) 
gives a general expression for the force exerted by radiation on a plasma with 
a given distribution function $\f(p)$. For a monochromatic beam of frequency 
$\omega$ propagating at angle $\ang$ with respect to the magnetic field, 
this expression gives the following momentum deposition rate,
\beq
    \frac{dP}{dt\,dV}=2\pi^2 r_e n F\xi\,\frac{\omega_B}{\omega^2}
           \left[\gamma_1\f(p_1)-\gamma_2\f(p_2)\right], \quad \omega\sin\ang\leq\omega_B,
\eeq
($dP/dt dV=0$ if $\omega\sin\ang>\omega_B$), where 
\beq
    p_{1,2}(\omega,\ang)=\frac{\omega_B}{\omega\sin^2\ang}
        \left(\cos\ang\mp\sqrt{1-\frac{\omega^2}{\omega_B^2}\sin^2\ang}\right),
\eeq
and the factor $\xi$ depends on the beam polarization,
\beq
       \xi=\left\{\begin{array}{ll}
                   1, &  \perp \\ 
              1-\displaystyle{\frac{\omega^2}{\omega_B^2}\sin^2\ang}, &  \parallel 
                   \end{array}
                   \right.
\eeq
The rate of momentum deposition by the beam (i.e. the exerted force)
per unit length along its trajectory is given by
\beq
   \frac{dP}{dt\,ds}=\frac{dP}{dt\,dV}\,A.
\eeq
We need to tabulate the force on a spatial grid, which is used to calculate the 
outflow dynamics at the next iteration. Therefore, we need to evaluate the net 
force applied to a given spatial cell.
The number of particles in the cell is $nV_c$ where $V_c$ is the cell volume,
and the force exerted by the beam {\it per particle} in the cell is given by
\beq
\label{eq:pc}
   \frac{d\FF_b}{ds}=\frac{1}{nV_c}\,\frac{dP}{dt\,ds}
   =2\pi^2\,\frac{r_e}{V_c}\,\frac{\omega_B\,\dE_b}{\omega^2}\,\xi
                    \left[\gamma_1\f(p_1)-\gamma_2\f(p_2)\right].
\eeq
In the transfer calculation, we track the photon trajectory using small steps 
$\delta s$, much smaller than the cells of the spatial grid.
To obtain the force $\FF_b(r,\theta,\zeta)$
applied by the beam in a given cell we integrate \Eq~(\ref{eq:pc})
along the photon path where it crosses the cell.
Note that a finer spatial grid implies a larger $d\FF_b/ds\propto V_c^{-1}$;
however, it also implies that the cell is less frequently visited by photons
in our Monte-Carlo simulation, and the photons spend a shorter time in the cell;
therefore, the final result does not depend on the grid.

In an axisymmetric magnetosphere, the spatial cells $(i,j)$ are tori of 
volume $V_{i,j}=2\pi r_i^2\sin\theta_j\,\Delta r\,\Delta\theta$. 
The net force applied per particle in a given cell $(i,j)$ is obtained by 
summing up the contributions $\FF_b(r_i,\theta_j,\zeta)$ from all simulated 
beams,
\begin{eqnarray}
\nonumber
  && \FF(r_i,\theta_j,\zeta)   =  \frac{\dN}{\KMC}\sum_{k=1}^{\KMC} 
       \frac{2\pi r_e \hbar}{V_{i,j}}  \\
  && \times \int_{{\rm cell}(i,j)} \frac{\omega_B}{\omega}
         \,\xi\,\left[\gamma_1\f(p_1)-\gamma_2\f(p_2)\right]\,ds.
\label{eq:FMC}
\end{eqnarray}
Note that the beam may cross a given cell multiple times in an opaque 
magnetosphere, and all crossings contribute to the path integral in 
\Eq~(\ref{eq:FMC}). 
The $e^\pm$ distribution function $\f(p)$ is determined by the parameter 
$\zeta$ (assuming a given pair multiplicity $\M$, see \Sect~5).

As a test, one can apply \Eq~(\ref{eq:FMC}) to the simplest case of a 
transparent outflow exposed to the central thermal radiation. Then the 
expected $\FF$ can be directly calculated using \Eq~(\ref{eq:FF2}) in 
Appendix~B. We analytically verified that in this case \Eq~(\ref{eq:FMC}) 
is reduced to \Eq~(\ref{eq:FF2}). 
We also tested our numerical code; it reproduced the analytical result.

\subsection{Results}

The obtained self-consistent solution for the outflow dynamics is shown 
in Figure~9. We show $p_+$ rather than our dynamical parameter $\zeta$,
because $p_+$ is closely related to the radiation emitted by 
the outflow. We find that only particles with the highest momenta 
$p\approx p_+$ resonantly scatter thermal photons in the relativistic 
zone $p_+>1$; the remaining, dominant part of the momentum 
distribution does not participate in scattering. 
The Lorentz factor of the scattering particles is 
\beq
  \gsc\approx (1+p_+^2)^{1/2}.
\eeq 

\begin{figure}
\epsscale{1.3}
\plotone{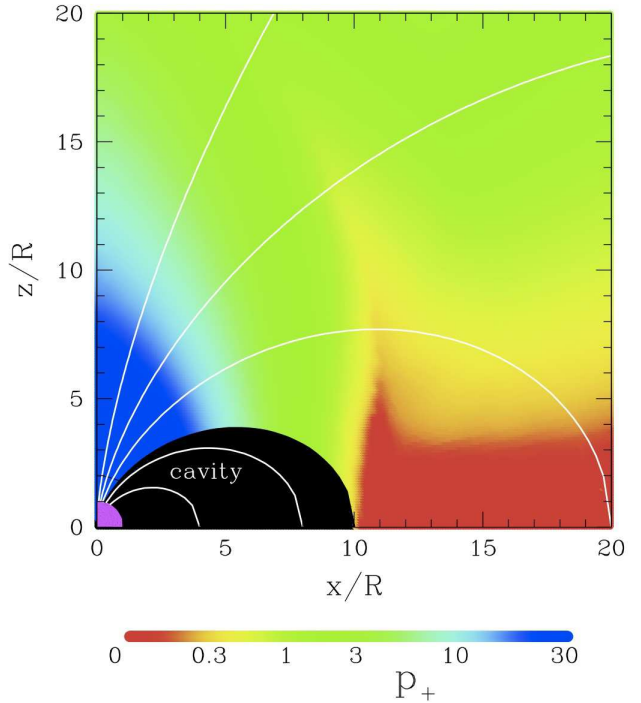}
\caption{Self-consistent outflow solution. Color shows the parameter $p_+$ 
of the waterbag distribution function. Observed hard X-rays (discussed 
in the accompanying paper Beloborodov (2013)), are produced by particles with 
$p\approx p_+$. The simulation assumed that the active j-bundle occupies 
the field lines with apex radii $\Rmax>10R$ (\Sect~1).
} 
\end{figure}

The solution $p_+(r,\theta)$ shown in Figure~9 was calculated assuming that 
the star has radius $R=10$~km, a uniform surface temperature 
$kT=0.3$~keV, and a moderately 
twisted dipole magnetic field with $B_{\rm pole}=10^{15}$~G and $\psi=0.3$; 
the multiplicity of the $e^\pm$ flow was fixed at $\M=200$.
The flow is injected at $r=2R$ with $p_+(2R)=100$. The choice of the
boundary condition is not important as we are interested in the flow behavior
outside $\sim 5R$, where the scattered photons avoid conversion to 
$e^\pm$ pairs and can escape, i.e. 
where the observed hard X-rays are produced (Beloborodov 2013).
For any reasonable boundary value $p_+(2R)$, the flow relaxes to the same
solution $p_+(r,\theta)$ outside a few stellar radii.
Remarkably, our simulations gave practically the same solution for 
a broad range of parameters $\psi$, $\M$, $T$ that is relevant for magnetars. 
This demonstrates a robust self-regulation mechanism; this important
feature is explained below (see also the accompanying paper).

There are two distinct zones in the outflow:

I. Non-relativistic zone $p_+\ll 1$, which is near the equatorial plane (red zone 
in Figure~9). This zone has a huge optical depth and scatters essentially all 
thermal photons that impinge on it from the star ($\sim 10$\% of the thermal
luminosity $\Lth$).
We will call this zone the ``equatorial reflector.''

II. Relativistic zone $p_+>1$ (blue to green in Figure~9). This zone is 
transparent for essentially all thermal photons flowing from the star, 
except for rare photons in the far Wien tail of the Planck spectrum, which
may be neglected. Basically, the outflow in this zone does not ``see'' the 
thermal radiation flowing from the star. 
It mainly interacts with (and is decelerated by) 
the quasi-thermal photons that flow from the equatorial equator.

It is easy to see why the outflow in the relativistic zone $\gamma_+\gg 1$ 
is regulated so that it interacts with a tiny fraction of the thermal photons 
around the magnetar. Scattering on average boosts the energy of a thermal 
photon $\hbar\omega$ by a factor comparable to $\gamma^2$, and hence 
the energy lost by the outflow per scattering is $\sim\gamma^2\hbar\omega$.  
If we imagine that each thermal photon is scattered once with an average 
blueshift of $\gamma^2$, the generated hard X-ray luminosity $\gamma^2\Lth$
would exceed the kinetic power of the outflow $L$, which is impossible.
The scattering rate must be kept low, just sufficient for the self-consistent  
gradual deceleration of the outflow.
The outflow ``vision'' of targets for scattering is controlled by
the resonance condition $\gamma(1-\beta\cos\ang)\omega=\omega_B$, 
where $\ang$ is the angle between the photon direction and the particle 
velocity $\vec{\beta}$. The scattering rate is greatly reduced if $\gamma$ 
is so low that the resonance condition is satisfied only for rare photons in 
the Wien tail of the thermal spectrum. The situation 
may be compared with the regulation of the nuclear burning rate
in the sun. The self-consistent temperature is sufficiently 
low so that fusion reactions occur only in the far tail of the Maxwell distribution;
as a result only rare particles participate in burning, and these particles 
are concentrated in a narrow interval of (large) momenta, a phenomenon 
known as the Gamow peak. Similarly, our self-consistent outflow moves 
sufficiently slow so that resonant scattering is only enabled between 
rare particles with the highest $p\approx p_+$ and 
rare thermal photons with the highest energies in the particle rest frame.
These lucky photons have the largest $\ang$ (gained after reflection from 
the equatorial reflector) and large $\hbar\omega\gg 2.7kT$.
The inward direction ($\cos\ang<0$) of the reflected photons gives them
particularly large blueshift in the outflow rest frame and hence reduces the 
energy requirement in the lab frame. This makes the reflected photons the 
dominant targets for 
scattering even though their density is much smaller than the density of 
photons flowing directly from the star. The reflected photons have a diluted 
quasi-thermal spectrum of temperature $T$. 

In essence, we observe in our simulations that the outflow moves fast enough 
to resonantly interact with reflected photons of energy 
$\hbar\omega\sim (7-10)kT$ 
(the low-density exponential tail of the thermal spectrum), and slow enough to 
not interact with the main peak of the reflected thermal spectrum 
$\hbar\omega\sim 3kT$. This condition, together with the resonance condition 
$\gamma(1-\beta\cos\ang)\omega=\omega_B$ and $\cos\ang\sim -0.5$, 
determines that the scattering plasma moves with Lorentz factor 
\beq
\label{eq:gsc}
     \gsc\approx \frac{m_ec^2}{10kT}\,\frac{B}{\BQ},  
\eeq
as long as $\gsc\gg 1$. In this regime, the number 
of target photons visible to an outflowing particle has a strong exponential
sensitivity to the particle momentum $p$. Essentially all scattering must be 
done by particles with the highest momenta in the distribution function, i.e. 
with $p\approx p_+$ for the waterbag distribution, as indeed observed
in our numerical simulation. Therefore, $\gsc$ is associated with $p_+$.
\Eq~(\ref{eq:gsc}) serves as a simple and reasonably accurate approximation 
to the exact numerical results shown in Figure~9. 
Its applicability is not limited to the specific simulation with 
its $B_{\rm pole}$, $T$, $\psi$, $\M$ --- the approximation 
works well for other magnetar parameters, because of the robust 
self-regulation effect described above.
This fact is further discussed and illustrated in Figure~2 in Beloborodov (2013). 

Besides \Eq~(\ref{eq:gsc}), the transfer problem is characterized by the 
position of the equatorial reflector. It is described by a simple 
formula, which can be used to scale the results shown in Figure~9 
to models with other parameters.
The formula is based on the following fact (see Appendix~B):
if a given active magnetic loop extends to the region where 
$\hbar\omega_B<20kT$, 
the central thermal radiation exerts a sufficiently strong 
drag on the outflow to bring it to rest at the top of the loop.
The region $\hbar\omega_B<20kT$ corresponds to $r>R_1$ where
\beq
\label{eq:R1}
   R_1\approx 80\, \left(\frac{\mum}{10^{33} {\rm~G~cm}^3}\right)^{1/3}
                                  \left(\frac{kT}{1~\rm keV}\right)^{-1/3}   {\rm ~km}.
\eeq
Here 
$\mum=R^3B_{\rm pole}/2$ is the magnetic dipole moment of the star.
\Eq~(\ref{eq:R1}) describes the position of the inner edge of the 
non-relativistic (red) zone; e.g. the model in Figure~9 has $R_1\approx 10R$.

It is instructive to compare the result of the full transfer calculation in Figure~9
with the simplest, optically thin two-fluid model in Figure~3.  
The relativistic zone $p_+>1$ remains practically transparent to 
thermal radiation. The key difference is the presence
of the opaque equatorial reflector.
The reflector weakly affects the spectrum of 
thermal photons supplied by the star, however it significantly changes their 
angular distribution in the magnetosphere. 
As a result, the radiation exerts a stronger drag on
the outflow and $p_+$ decreases faster along the magnetic field lines.
In Figure~3,  
$\gamma_+\approx p_+\gg 1$ remains huge near the magnetic axis --- the 
central radiation is unable to decelerate the plasma because the photons flow 
from behind and have small angles $\ang$ with respect to the plasma velocity.
In Figure~9, the equatorial reflector supplies photons with large $\ang$,
which efficiently decelerate the outflow, according to \Eq~(\ref{eq:gsc}). 

The upscattered photons of energy $E\sim \gamma^2\Etarget$ are beamed 
along the relativistic outflow. Therefore, they become unable to decelerate the 
plasma, even though they can scatter multiple times before escaping.
The outflow significantly loses energy when it scatters a photon 
propagating at a large angle $\ang$ with respect to the outflow velocity; 
only in this case
the scattering boosts the photon energy by the factor of $\sim \gsc^2\gg 1$.
After the scattering, the photon angle is reduced to $\ang\sim\gsc^{-1}$,
and its subsequent scatterings have a small effect on 
the outflow dynamics. The beamed radiation initially moves together with the 
plasma and then escapes. 
Our transfer simulations include all scattering events,
however practically the same $p_+(r,\theta)$ would be obtained if only 
single scattering were allowed in the relativistic zone $p_+\gg 1$.


\section{Conclusions}
\label{sec:3}

This paper examined the behavior of the relativistic plasma
created by $e^\pm$ discharge around magnetars. 
Plasma circulation in the magnetosphere is schematically shown in Figure~2.
We focused on large magnetic loops, 
which must be heavily loaded with $e^\pm$ pairs. 
The plasma momentum is controlled by the radiation field around the star, 
which interacts with $e^+$ and $e^-$ via resonant scattering.
We developed a method to calculate radiative transfer in the self-consistently
moving plasma and obtained the solution for the $e^\pm$ flow (Figure~9). 
The solution is a strong attractor --- the behavior of the plasma outside a few 
stellar radii does not depend on 
how the plasma is injected near the star and its initial Lorentz factor $\gamma_0$,
as long as $\gamma_0\gg 30$ ($\gamma_0\simgt 10^3$ is 
expected, which corresponds to the discharge voltage $\Phi_e\sim 10^9$~V).
The $e^\pm$ flow shows the following features:

(1) The relativistic flow scatters radiation with a well defined 
Lorentz factor $\gsc$ given in \Eq~(\ref{eq:gsc});
$\gsc$ decreases proportionally to $B$ along the magnetic field lines. 

(2) The relativistic flow remains transparent to thermal photons emitted
by the star until it decelerates to non-relativistic momenta $p<1$. 
The non-relativistic zone is opaque to the star radiation and forms the 
``equatorial reflector'' (red zone in Figure~9).

(3) The energy lost by the decelerating flow is converted to hard X-rays. 
The resulting emission is calculated and compared with observations 
in the accompanying paper (Beloborodov 2013).

(4) The plasma is nearly neutral, $n_+\approx n_-$, and carries the electric 
current $\bj=(c/4\pi)\nabla\times\bB$ by adjusting the particle velocities.
In large magnetic loops (extending to the region of $B\simlt 10^{13}$~G) the 
plasma has a high $e^\pm$ multiplicity $\M\sim 10^2$, and both electrons and 
positrons outflow from the neutron star, with a small separation in the velocity 
space. This separation is sustained by a 
modest electric field induced along the magnetic field lines. 

(5) The enforced electric current and radiative drag together create a 
configuration that is prone to two-stream instability, 
which is expected to generate low-frequency radiation. 
The mechanism of initiating the two-stream instability is unique to magnetars, 
explaining their special radio emission and possibly optical/UV emission.
 
\acknowledgments
This work was supported by NASA grants NNX-10-AI72G and 
NNX-13-AI34G. I thank R. Hasco\"et for comments on the manuscript.


\begin{appendix}


 \section{Resonant scattering}

Resonant scattering plays a significant role in ordinary pulsars
(e.g. Kardashev, Mitrofanov, \& Novikov 1984; Daugherty \& Harding 1989; 
Sturner 1995; Lyubarsky \& Petrova 2000). It is also the dominant radiative 
process in magnetar magnetospheres, which governs the radiative transfer 
calculated in this paper.
Below we summarize basics of resonant scattering, write down the cross section,
the optical depth of the $e^\pm$ flow, and the radiative drag force 
that are used in our numerical simulations. 

Photons scattered in the region $B>10^{13}$~G are immediately absorbed 
(Beloborodov 2013. Interesting radiative transfer occurs outside this region, 
where $B\ll\BQ$. In particular, the observed hard X-rays of energy up to several 
MeV are radiated where $B\ll\BQ$. Electron recoil is small for resonant scattering 
in such relatively weak fields, and the scattering cross section is particularly simple.

\subsection{Scattering cross section}
\label{sec:cs}

In classical language, the electro-magnetic wave (photon) with frequency
$\omega_B=eB/m_ec$ resonates with the Larmor rotation of electron. 
Then the wave strongly accelerates the charged particle and generates 
scattered radiation. The corresponding cross section is largest for 
waves with the right-hand circular polarization $\epol=\ee_-$ 
that matches the electron Larmor rotation.
Here $\ee_-=2^{-1/2}(\ee_x-i\ee_y)$ and $\{\ee_x$, $\ee_y$, $\ee_z\}$ 
is a Cartesian basis with the $z$-axis anti-parallel to $\bB$.
For a wave with an arbitrary polarization vector $\epol$, only the projection
of $\epol$ on $\ee_-$ is responsible for the resonance, and the cross 
section is reduced by the factor $|{\epol}^*\cdot\ee_-|^2$, where
${\epol}^*$ is the complex conjugate of $\epol$.
In quantum language, the resonance occurs because the photon energy 
matches the energy $\hbar\omega_B$ needed for the electron transition 
from the ground Landau state to the first excited state. 
The resonance has a finite width. It equals the natural width of the 
cylcotron line $\Gamma$, which corresponds to the lifetime of the excited 
electron to spontaneous transition back to the ground Landau state.

For an electron at rest, the differential cross section for photon 
scattering into solid angle $d\Omegasc$ is given by (Canuto et al. 1971; 
Ventura 1979),
\beq
\label{cs1}
   \frac{d\sigma}{d\Omegasc}=r_e^2\, 
        \frac{\omega^2}{(\omega-\omega_B)^2+(\Gamma/2)^2}
          \,|{\epol}^*\cdot\ee_-|^2\,|{\epolsc}^*\cdot\ee_-|^2,
\eeq
where $r_e=e^2/m_ec^2$, $\omega$ is the photon frequency,
$\epol$ and $\epolsc$ are the polarization vectors of the photon 
before and after scattering.
Equation~(\ref{cs1}) retains only the resonance peak of the cross section
and neglects the non-resonant part.
Positron cross section is described by the same equation except that 
$\ee_-$ is replaced by $\ee_+=2^{-1/2}(\ee_x+i\ee_y)$ 
(positrons gyrate in the opposite sense).
The cross section can also be derived in the framework of quantum 
electrodynamics (Herold 1979; Daugherty \& Harding 1986). For  
$\hbar\omega_B\ll m_ec^2$ (which corresponds to $B\ll\BQ$) the result is 
reduced to \Eq~(\ref{cs1}).

The resonance line is very narrow, $\Gamma/\omega_B=(4/3)\alf(B/\BQ)\ll 1$, 
where $\alf=e^2/\hbar c=1/137$ (Daugherty \& Ventura 1978; Herold et al. 1982),
and the resonance factor $[(\omega-\omega_B)^2+(\Gamma/2)^2]^{-1}$ is well 
approximated by the delta-function $2\pi\Gamma^{-1}\delta(\omega-\omega_B)$. 
Then the cross section may be written as 
\beq
\label{cs}
  \frac{d\sigma}{d\musc}=2\pi\,\frac{d\sigma}{d\Omegasc}
        =3\pi^2 r_e c\,\delta(\omega-\omega_B)
        \,|{\epol}^*\cdot\ee_\pm|^2\,|{\epolsc}^*\cdot\ee_\pm|^2,
\eeq
where $+/-$ correspond to scattering by positron/electron,
$\musc=\cos\angsc$, and $\angsc$ is the angle 
of the scattered photon with respect to the magnetic field $\bB$. 
The distribution of scattered photons is axially symmetric about $\bB$;
this fact has been used in \Eq~(\ref{cs}).

Two polarization states (eigen modes) exist for the photon. 
They are controlled by the dielectric tensor of the magnetosphere. 
In the considered region $r<100R$, the dielectric tensor for photons 
of interest (X-rays) is dominated by the magnetic vacuum polarization effect 
(e.g. Beresteskii et al. 1982); the plasma 
contribution to the dielectric tensor is much smaller and may be neglected.
Magnetic vacuum defines two linearly polarized eigen modes for 
electromagnetic waves:
$\ee_\perp$ which is perpendicular to the $(\k,\bB)$ plane, and 
$\ee_\parallel$ which is parallel to the $(\k,\bB)$ plane
(here $\k$ is the photon wave vector and $\ee$ shows the direction of 
the electric field in the wave). 
The $\perp$ and $\parallel$ modes are also called E-mode and O-mode, 
respectively. Their refraction indices are
\beq
  N_\perp=1+\frac{2\alf}{45\pi} \left(\frac{B}{\BQ}\right)^2\sin^2\ang, \qquad
  N_\parallel=1+\frac{7\alf}{90\pi} \left(\frac{B}{\BQ}\right)^2\sin^2\ang,
\eeq
where $\ang$ is the photon angle with respect to $\bB$. 
The two modes have slightly different propagation speeds $c/N$ and
therefore they adiabatically track, i.e. the photon propagating through 
the curved magnetic field preserves its polarization state. 
The adiabaticity condition reads $kl_B(N_\parallel-N_\perp)\gg 1$ where 
$l_B\sim r$ is the characteristic scale of the spatial variation of $\bB$
(see e.g. Fern\'andez \& Davis 2011 for a detailed discussion). 
This condition is satisfied for X-rays in the considered region of the 
magnetosphere where scattering occurs. Thus, in our transfer problem, 
the photon can switch its polarization state only in a scattering event. 

As the photon can be in either polarization state,
calculation of radiative transfer involves four scattering processes
$\perp\rightarrow\perp$, $\perp\rightarrow\parallel$, 
$\parallel\rightarrow\perp$, and $\parallel\rightarrow\parallel$.
The corresponding cross sections are given by \Eq~(\ref{cs})
with $|\ee_\perp^*\cdot\ee_\pm|^2=|{\ee_\perp^\prime}^*\cdot\ee_\pm|^2=1/2$,
$|\ee_\parallel^*\cdot\ee_\pm|^2=\mu^2/2$, and 
$|{\ee_\parallel^\prime}^*\cdot\ee_\pm|^2={\mu^\prime}^2/2$.
Note that the cross sections of electron and positron are equal, as 
$|\epol^*\cdot\ee_-|^2=|\epol^*\cdot\ee_+|^2$ for any linear polarization 
$\epol$.

Equation~(\ref{cs}) describes the cross section of electron (or positron) 
at rest. In our transfer problem, the particles are moving along $\bB$ 
and the above equations should be used in the rest frame of the particle. 
Note that the polarization states and $\omega_B$
are invariant under Lorentz boosts along $\bB$.
Consider a photon with energy $\hbar\omega$ and propagation angle $\ang$
with respect to $\bB$, in the $\perp$ or $\parallel$ polarization state.
We are interested in its scattering by an electron (or positron)
that moves with velocity $\beta c$ along the magnetic field line.
The total cross section in the lab frame may be obtained by integrating 
the differential cross section in the electron rest frame and then 
multiplying the result by $1-\beta\cos\ang$,
\beq
\label{eq:sigtot}
  \sigtot = 2\pi^2\,r_e c\,\xi\,\delta(\omegac-\omega_B)\,(1-\beta\mu),
\eeq
where 
\beq
\label{omc}
   \omegac=\gamma(1-\beta\mu)\,\omega,  
\eeq
is the photon frequency in the electron rest frame and $\mu=\cos\ang$. 
The factor $\xi$ depends on the photon polarization and is given by
\beq
\label{xi}
   \xi\equiv\left\{\begin{array}{ll}
                   1, &  \perp \\ 
              \muc^2, &  \parallel 
                   \end{array}
            \right\},
  \qquad \muc=\frac{\mu-\beta}{1-\beta\mu},
\eeq
where $\muc=\cos\angc$ and $\angc$ is the photon angle with 
respect to $\bB$ in the electron rest frame. 
The cross section $\sigtot$ is summed over the final polarization states.

The outcome of the scattering may be either $\perp$ or $\parallel$ photon
propagating at angle $\angc^\prime$ in the electron frame.
The probability distribution for $\mucsc=\cos\angc^\prime$ is given by 
\beq
   P(\mucsc)=\frac{1}{\sigtot}\,\frac{d\sigma}{d\mucsc}=\frac{3}{8}\,\xisc.
\eeq
where $\xisc=1$ or $\muc^{\prime 2}$, depending on the final polarization 
state.
The integral $\int P(\mucsc)\,d\mucsc$ equals 3/4 for the $\perp$ state and 
1/4 for the $\parallel$ state. Thus, 3/4 of scattering events produce $\perp$ 
photons with uniform distribution $P(\mucsc)=const$ and the remaining 1/4
of scattering events produce $\parallel$ photons with 
$P(\mucsc)\propto\muc^{\prime 2}$.
The distribution of the final photons over angle and polarization does not 
depend on the initial state of the photon before scattering.

The standard description of resonant scattering summarized above assumes 
the transition between the ground and first excited Landau levels, which has 
the largest cross section. Transitions to higher levels are neglected in our transfer 
calculations.

\subsection{Opacity}
\label{sec:OD}

Optical depth of a relativistic plasma to resonant scattering was discussed 
previously in detail (e.g. Fern\'andez \& Thompson 2007 and refs. therein). 
Here we write down relevant equations and introduce notation that is
used below in the discussion of radiative drag.
Consider a photon of energy $\hbar\omega$
that propagates through $e^\pm$ plasma with density $n=n_++n_-$.
The plasma particles move along $\bB$ with 
momentum distribution $\f(p)$ that is normalized to unity
\beq
   \int \f(p)\,dp = 1.
\eeq
The optical depth $d\tau$ seen by the photon as it propagates distance $ds$ 
is given by
\beq
\label{eq:dtau}
   \frac{d\tau}{ds}=n\int \sigtot\,\f(p)\,dp.
\eeq
Assuming that the magnetosphere has a small optical depth to non-resonant
scattering (Appendix~C), we include only resonant scattering 
and use \Eq~(\ref{eq:sigtot}) for $\sigtot$.
Integration over $p$ may be carried out using the identity,
\beq
\label{eq:delta}
   \delta(\omegac-\omega_B)=\sum_{k} \frac{\delta(p-p_k)}{|d\omegac/dp|},
\eeq
where
\beq
  \frac{d\omegac}{dp}=\frac{d}{dp}\left(\gamma-p\mu\right)\,\omega
      =(\beta-\mu)\omega,
\eeq
and $p_k$ are all possible solutions of equation $\omegac(p)=\omega_B$.
The delta-functions $\delta(p-p_k)$ express the fact that the photon is
scattered by electrons or positrons with momenta $p_k$ for which
the resonance condition is met.
The relation $\omegac\sin\angc=\omega\sin\ang$ together with 
the resonance condition $\omegac=\omega_B$ determines 
$\sin\angc=(\omega/\omega_B)\sin\ang$ and leaves two 
possibilities for $\muc=\cos\angc$,
\beq
\label{eq:muc}
  \muc=\pm\left(1-\frac{\omega^2}{\omega_B^2}\sin^2\ang\right)^{1/2}.
\eeq
Angle $\angc$ exists (i.e. the resonance is in principle possible) 
for photons that satisfy the condition $\omega\sin\ang\leq\omega_B$.
Then \Eq~(\ref{eq:muc}) defines two electron velocities $\beta_{1,2}$,
which may be found from the Doppler transformation of the photon angle, 
$\muc=(\mu-\beta)/(1-\beta\mu)$. It yields,
\beq
  \beta=\frac{\mu-\muc}{1-\mu\muc}, \qquad 
  \beta_{1,2}=\frac{\mu\mp|\muc|}{1\mp\mu|\muc|}.
\eeq
The corresponding Lorentz factors $\gamma=(1-\beta^2)^{-1/2}$ and
dimensionless momenta $p=\gamma\beta$ are 
\beq
\label{eq:gam}
   \gamma_{1,2}=\frac{1\mp\mu|\muc|}{\sin\ang\sin\angc}, \qquad
   p_{1,2}=\frac{\mu\mp|\muc|}{\sin\ang\sin\angc}.
\eeq

Substitution of \Eqs~(\ref{eq:sigtot}) and (\ref{eq:delta}) to
\Eq~(\ref{eq:dtau}) and integration over $p$ gives 
\beq
\label{eq:dtau1}
  \frac{d\tau}{ds}= 2\pi^2\,r_e\,\frac{c}{\omega}\,\frac{\xi}{|\muc|}\, 
     n\,\left[\f(p_1)+\f(p_2)\right].
\eeq
Here $\xi=1$ for $\perp$ photons and $\xi=\muc^2$ for $\parallel$ photons 
(\Eq~\ref{xi}).

\subsection{Radiative drag force}

Consider now a radiation field with intensity $I(\omega,\k)$ in a given
polarization state, $\perp$ or $\parallel$.
As a result of scattering, radiation exerts a force on the $e^\pm$ plasma.
We are interested in the component of this force along the magnetic field.
The force applied to unit volume of the plasma is given by
\beq
\label{eq:dP}
  \frac{dP}{dt\,dV}=\int d\Omega \int d\omega \int dp \;
     \frac{I(\omega,\k)}{\hbar\omega}\,n\,\f(p)\,\sigtot\,\Delta P.
\eeq
Here $\int d\Omega$ is the solid-angle integration over photon directions 
$\nn=\k/k$, and $\Delta P$ is the average momentum (per scattering) 
passed to an electron or positron with Lorentz factor $\gamma$ by a photon 
$(\omega,\k)$. 
  In the electron rest frame, $\Delta\Pc$ equals the photon momentum along 
$\bB$, as its average momentum after scattering vanishes
(resonant scattering is symmetric in the electron frame when 
$\hbar\omega_B\ll m_ec^2$).
Thus, $\Delta\Pc =\muc\,\hbar\omegac/c$ and the Lorentz 
transformation of the four-momentum vector to the lab frame gives
\beq
\label{eq:DeltaP}
  \Delta P=\gamma\,\muc\frac{\hbar\omega_B}{c},
\eeq
where we used the condition $\omegac=\omega_B$ since we consider only 
resonant scattering.

Radiation is described by two intensities $I_\perp(\omega,\k)$
and $I_\parallel(\omega,\k)$ in the two polarization states. 
They scatter with cross sections $\sigtot$ that differ by the factor 
of $\muc^2$ (see \Eqs~(\ref{eq:sigtot}) and (\ref{xi})). 
Substituting \Eq~(\ref{eq:sigtot}) to \Eq~(\ref{eq:dP}) and
taking the sum over polarizations, one finds
the net force exerted by $I_\perp$ and $I_\parallel$,
\beq
\label{eq:dP1}
  \frac{dP}{dt\,dV}=\int d\Omega \int d\omega \int dp \;
     \frac{(I_\perp+\muc^2I_\parallel)}{\hbar\omega}
     \,n\,\f(p)\,2\pi^2 r_ec\,\delta(\omegac-\omega_B) (1-\beta\mu)\,\Delta P.
\eeq
Integration over $dp$ similar to that in \Sect~\ref{sec:OD} gives
\beq
\label{eq:dP2}
  \frac{dP}{dt\,dV}=\int d\Omega \int_0^{\omega_B\sin\ang} d\omega \;
 2\pi^2 r_e\,\frac{\omega_B}{\omega^2}\left(I_\perp+\muc^2I_\parallel\right)
  n \left[\gamma_1\f(p_1)-\gamma_2\f(p_2)\right],
\eeq
where $p_{1,2}(\omega,\ang)$ and $\gamma_{1,2}$ are given by 
\Eqs~(\ref{eq:gam}) and (\ref{eq:muc}).
The upper limit in the integral over $\omega$ takes into account
that only photons with $\omega\leq\omega_B\sin\ang$ may be resonantly 
scattered (\Sect~\ref{sec:OD}).
Equation~(\ref{eq:dP2}) is useful because it shows the contribution 
of each photon $(\omega,\k)$ (in the $\perp$ or $\parallel$ polarization state) 
to the drag force. It can be used even if the radiation intensity 
is not known in advance and needs to be found self-consistently with the 
plasma dynamics (\Sect~6). 

An alternative way of simplifying \Eq~(\ref{eq:dP1})
is to first integrate over $\omega$, which gives 
\beq
\label{eq:dP3}
  \frac{dP}{dt\,dV}=\int d\Omega \int dp \; 2\pi^2 r_e\,
     (I_\perp+\muc^2I_\parallel)\,n\,\f(p)\,\gamma(\mu-\beta),
\eeq
where $I_\perp$ and $I_\parallel$ are evaluated at 
$\omega=\gamma^{-1}(1-\beta\mu)^{-1}\omega_B$. Equation~(\ref{eq:dP3}) is 
convenient to use when the radiation intensity is known.

Note that Equations~(\ref{eq:dP2}) and (\ref{eq:dP3}) are valid only 
where $B\ll\BQ$. Near the star, where the field is stronger, the drag force is
modified  (Baring et al. 2011; Beloborodov 2013).
In this paper, we do not need the strong-field corrections, as radiative 
transfer occurs in the region of $B\ll \BQ$ (where the scattered photons avoid the 
quick conversion to $e^\pm$ pairs); use of the full relativistic cross section would 
be an unnecessary complication.


\section{Optically thin outflow}
\label{sec:thin}

Consider a magnetosphere that is optically thin to resonant scattering, so 
that intensity $I$ is dominated by the unscattered radiation from the star. 
We will assume that the neutron star emits approximately blackbody radiation 
with the $\perp$ polarization (the $\perp$ photons dominate the surface radiation
because they have a larger free path below the surface, e.g. 
Silant'ev \& Iakovlev 1980). Then
\beq
  I_\perp=\frac{\hbar\omega^3}{8\pi^3c^2[\exp(\hbar\omega/\kB T)-1]},  
  \qquad  I_\parallel=0,
\eeq
and we deal with the outflow dynamics in the known radiation field. 
This case was studied in detail in previous work (e.g. Sturner 1995).

\subsection{Scattering rate for one particle}

Consider an electron (or positron) located at $r,\theta$ and moving outward 
with Lorentz factor $\gamma=(1-\beta^2)^{-1/2}$ along a magnetic field line. 
The number of photons scattered by the electron per unit time is 
\beq
\label{eq:dNsc1}
   \dNsc=\int d\Omega \int d\omega\, \frac{I(\omega,\nn)}{\hbar\omega}\,\sigtot
        =\frac{2\pi^2 r_ec}{\gamma} \int \frac{I(\omegares[\nn],\nn)}
                                       {\hbar\omegares}\,d\Omega,
\eeq
where we substituted $\sigtot$ (eq.~\ref{eq:sigtot}). The resonant frequency
depends on the photon direction, 
$\omegares=\gamma^{-1}(1-{\mathbf\beta}\cdot\nn)^{-1}\omega_B$
where $\nn=\k/k$ is the unit vector corresponding to $d\Omega$.

At large radii $r\gg R$ all photons at a given location $\br$ have 
approximately the same direction $\nn\parallel\br$.
The angle between the stellar photons and the particle velocity, $\ang$,
is given by $\mu=\cos\ang\approx B_r/B$ (assuming $\ang>R/r$). Then
integration over $d\Omega$ in \Eq~(\ref{eq:dNsc1}) is reduced to 
multiplication by $\Delta\Omega\approx\pi(R/r)^2$, the solid angle 
subtended by the star when viewed from radius $r$. This gives,
\beq
\label{eq:dNsc2}
   \dNsc=\frac{2\pi^3 r_ec}{x^2\gamma}\,\frac{I(\omegares)}{\hbar\omegares}
   \qquad \omegares=\frac{\omega_B}{\gamma(1-\beta\mu)},
\eeq
where $x=r/R$. 
 It is instructive to write $\dNsc$ in the following form,
\beq
\label{eq:dNsc3}
  \dNsc=\frac{\alf\,\Theta^2\,c}{4x^2\gamma\,\lbar}\,\frac{g(y)}{y},
\eeq
where $\lbar=\hbar/m_ec$, $\Theta=\kB T/m_ec^2\approx 10^{-3}$ and $g(y)$ 
is the dimensionless Planck function evaluated at the frequency 
$\omega_{\rm res}=\gamma^{-1}(1-\beta\mu)^{-1}\omega_B$,
\beq
\label{eq:g}
   g(y)=\frac{y^3}{e^y-1}, \qquad
   y=\frac{\hbar\omega_{\rm res}}{\kB T}=\frac{b}{\gamma(1-\beta\mu)\Theta},
\eeq
where $b=B/\BQ$.

\subsection{Drag force exerted on one particle}

The drag force applied by the central blackbody radiation to the electron 
is $\F=\dNsc\Delta P$ where $\Delta P$ is given by \Eq~(\ref{eq:DeltaP}).
This yields,
\beq
\label{eq:f}
  \F(\beta)=\frac{\alf^2}{4x^2}\frac{m_ec^2}{r_e}\,\Theta^3
  \gamma\,g(y)\,(\beta_\star-\beta),
\eeq
where $\beta_\star=\mu$.
Force $\F$ vanishes if $\beta=\beta_\star$; in this case the radiation
flux measured in the rest frame of the particle is perpendicular to $\bB$
and cannot accelerate or decelerate it.
In a weakly twisted magnetosphere, the magnetic field in the outer
corona is approximately dipole. 
Then the saturation velocity $\beta_\star$ at a point $r,\theta$ (spherical
coordinates) depends only on $\theta$ and is given by
\beq
\label{eq:sat}
  \beta_\star=\mu=\frac{B_r}{B}=\frac{2\cos\theta}{(1+3\cos^2\theta)^{1/2}},
 \qquad p_\star=\frac{2\cos\theta}{\sin\theta}.
\eeq
The radiative force always pushes the particle toward $p=p_\star$.
This effect may be measured by the ``drag coefficient,''
\beq
\label{eq:D}
   \D\equiv \frac{r}{c}\,\frac{1}{p}\,
  \frac{dp}{dt}.
\eeq
Consider an electron (or positron) with momentum 
$p\approx p_\star=\gamma_\star\beta_\star$. A small deviation $p-p_\star$ 
causes drag $\D\propto p-p_\star$, which may be written as
\beq
\label{eq:Dstar1}
   \D=\D_\star\left(1-\frac{p}{p_\star}\right), 
\eeq
where
\beq
\label{eq:Dstar}
   \D_\star=\frac{\alf^2}{4}\frac{R}{r_e}
            \frac{\Theta^3g(y_\star)}{x\,\gamma_\star^2}
     \approx \frac{4\times 10^4}{x} \frac{g(y_\star)}{\gamma_\star^2}
             \left(\frac{\kB T}{0.5\rm ~keV}\right)^3.
\eeq
Here $y_\star=b\gamma_\star/\Theta$ corresponds to photons that are
resonantly scattered by the electron with $p\approx p_\star$.
The momentum $p_\star$ is a strong attractor if $\D_\star\gg 1$.
The value of $\D_\star$ is sensitive to $y_\star$.
In particular, in the equatorial plane, we have $\gamma_\star=1$ and
\beq
  y_\star
       \approx 1.6\times 10^4 \left(\frac{B_{\rm pole}}{10^{15}\rm ~G}\right)
                   \left(\frac{\kB T}{0.5\rm ~keV}\right)^{-1}
         \left(\frac{r}{R}\right)^{-3},  \qquad (\theta=\pi/2),
\eeq
where $B_{\rm pole}$ is the dipole field at the magnetic pole.
The condition $\D_\star>1$ corresponds to $y_\star\simlt 20$. 
This implies that the 
$e^\pm$ flow on magnetic field lines extending far from the star 
is stopped by the radiative drag in the equatorial plane.
For typical magnetar parameters, the flow stops on field lines with 
$\Rmax\simgt 10R$.

\subsection{Optical depth in the single-fluid approximation}

The single-fluid flow has a distribution function $\f(p^\prime)=\delta(p^\prime-p)$
where $p(r,\theta)$ is the flow momentum.
Then any photon of energy $\hbar\omega$ may only be scattered on the 
infinitesimally thin resonance surface defined by 
$\gamma(1-\beta\mu)\omega=\omega_B$, where 
$\mu=\cos\ang$ describes the photon angle relative to the flow velocity.
The optical depth of the resonant surface
may be obtained from \Eq~(\ref{eq:dtau}), which gives
\beq
   \frac{d\tau}{ds}=2\pi^2 r_e c\,n\,\xi\, (1-\beta\mu)\delta(\omegac-\omega_B),
\eeq
where $\omegac=\gamma(1-\beta\mu)\omega$. One can use the identity, 
\beq
  \delta(\omegac-\omega_B)=\sum_k \frac{\delta(s-s_k)}
                          {\left|\frac{d}{ds}(\omegac-\omega_B)\right|}.
\eeq
The location $s_k$ on the photon trajectory is where the photon crosses 
the resonant surface.
Performing the integration over $s$ along the photon trajectory, one finds 
the optical depth for one crossing of the resonant surface,
\beq
   \tau=\frac{2\pi^2 r_e c\,n\,\xi\,(1-\beta\mu)}
           {\left|\frac{d}{ds}(\omegac-\omega_B)\right|}.
\eeq
If we specialize to the case of central photons emitted by the neutron 
star with the $\perp$ polarization,\footnote{
      The neutron-star radiation is dominated by the $\perp$ polarization
      (Silant'ev \& Iakovlev 1980).
      In addition, for the outflow with the equilibrium momentum $p_\star$ 
      considered below, the scattering of $\parallel$ photons (even if 
      they were included) would be suppressed. In the outflow rest frame, the 
      photons move perpendicular to $\bB$, and the resonant cross section for 
      the $\parallel$ polarization mode vanishes.} 
then $\xi=1$ and $\mu=B_r/B$.
For a moderately twisted dipole magnetosphere, the electric current density 
is given by $j\approx c\psi\,B/4\pi \Rmax$ (Beloborodov 2009), and 
\beq
    n=\M\,\frac{j}{ev}\approx\frac{\M \psi B}{4\pi e \beta \Rmax}.
\eeq
Note that $n$ is small on field lines with a large $\Rmax$, which implies
a low optical depth near the axis; this fact was also emphasized by  
Thompson et al. (2002) who used a self-similar twist model.

The expression for the optical depth becomes particularly simple if the outflow 
has the equilibrium momentum $p=p_\star$. Then $d\omegac/ds=0$, i.e. 
$\omegac$ remains constant along the radial ray through the outflow.
This fact can be derived by noting that the Doppler factor 
$\gamma(1-\beta\mu)=\gamma_\star^{-1}$ is 
a function of $\theta$ only --- it does not depend on $s=r$ for the approximately
 dipole magnetosphere. It is also easy to see that 
$d\omega_B/ds=-3\omega_B/r$, and we find
\beq
\label{eq:tau_dip}
  \tau=\frac{\pi}{12}\,\M\psi\,\frac{\sin^4\theta}
                                    {\cos\theta(1+3\cos^2\theta)^{1/2}}.
\eeq

The single-fluid model with $p=p_\star$ may approximate the outflow only 
sufficiently close to the equatorial plane where $1-\beta_\star\simgt\M^{-1}$.
Nevertheless, the approximate \Eq~(\ref{eq:tau_dip}) shows a general feature: 
the optical depth seen by the central photons is dramatically increased 
toward the equatorial plane ($\tau\propto \sec\theta$) and dramatically reduced 
toward the axis ($\tau\propto\sin^4\theta$). As a result, a distant observer can 
see the unscattered radiation from the neutron-star surface when the line of 
sight is within a moderate angle $\theta<\pi/4$ from the polar axis. 
This feature becomes even more pronounced in the full radiative transfer 
problem where the relativistic outflow is decelerated by the reflected
radiation from the outer corona. Then scattering of the central radiation 
is negligible in the entire relativistic zone of the outflow.

\subsection{Drag force exerted on a plasma with a broad distribution function}

\Eq~(\ref{eq:f}) describes the drag force exerted by the central thermal 
radiation on a particle with a given momentum $p=\gamma\beta$. One can 
also consider a collection of particles with a momentum distribution $\f(p)$
and derive the average force per particle $\FF=n^{-1}(dP/dVdt)$. 
\Eq~(\ref{eq:dP3}) gives
\beq
\label{eq:FF2}
   \FF=\frac{r_e\hbar}{4 c^2}\left(\frac{R}{r}\right)^2
       \omega_B^3 \int_{-\infty}^\infty \frac{(\mu-\beta)\,\f(p)}
         {\gamma^2(1-\beta\mu)^3(\exp y-1)}\,dp.
\eeq
where $y$ is given in \Eq~(\ref{eq:g}). The same result is obtained by averaging 
the force $\F$ given by \Eq~(\ref{eq:f}), $\FF=\int \F(p) \f(p)\,dp$. 

\Eq~(\ref{eq:FF2}) simplifies when the plasma is described by the waterbag
distribution function $\f$ (\Sect~5.1); it
leads to a straighforward 
calculation of the flow dynamics in the central radiation field.
We use this simple outflow model as the first trial to initiate the iterations
that converge to the solution shown in Figure~9. In the final solution, the drag 
exerted by the central radiation turns out negligible in the relativistic zone;
instead, the outflow deceleration is controlled by the radiation streaming from
the equatorial reflector, as discussed in \Sect~6.
Then the force $\FF$ derived in this section may be of interest only in the 
non-relativistic zone.


\section{Non-resonant scattering}

Non-resonant scattering is not limited by the resonance condition, and hence 
many more photons can participate in scattering, although with a smaller 
cross section. Below we discuss the effect of non-resonant scattering on 
the dynamics of $e^\pm$ flow around magnetars.

Non-resonant scattering occurs mainly with photons in the Wien peak of 
the thermal radiation flowing directly from the neutron star, which  
dominates the photon density around the star.
Relativistic particles see the thermal photons (of typical energy $E\sim 3kT$) 
blueshifted as  $\tilde{E}=\gamma(1-\beta\mu)E$ where 
$\mu=\cos\ang$ 
describes the photon direction relative to the particle velocity in the lab frame. 
Sufficiently far from the star (where $R^2/r^2<1-B_r/B$) the radiation can 
be approximated as a narrow radial beam; then 
$\mu=B_r/B$. In general, $\mu$ is a function of the particle position 
$r,\theta,\phi$ in the magnetosphere. For an approximately dipole field,
$\mu=2\cos\theta\, (1+3\cos^2\theta)^{-1/2}$ is a function of the polar angle
$\theta$ only.
 
Magnetic field strongly affects the non-resonant scattering cross section 
if $\tilde{E} < \hbar\omega_B=b m_ec^2$. If the electron is relativistic,
the target photons are aberrated in the electron rest frame,
$\cos\angc=\muc=(\mu-\beta)/(1-\beta\mu)$.
In the limit $\beta\rightarrow 1$, even photons with the $\parallel$ 
polarization have electric fields almost perpendicular to $\bB$, which 
makes their scattering inefficient.
For photons with $\tilde{E} \ll \hbar\omega_B$,
the non-resonant scattering cross section is given by
(e.g. Canuto et al. 1971) 
\beq 
\label{eq:sig_nonres}
  \frac{\sigma_\parallel}{\sT}\approx \left(\frac{\tilde{E}}{\hbar\omega_B}\right)^2
      +\frac{\sin^2\angc}{2},   \qquad
   \frac{\sigma_\perp}{\sT}\approx \left(\frac{\tilde{E}}{\hbar\omega_B}\right)^2,
   \qquad \Ec\ll\hbar\omega_B.
\eeq
We assume $\Ec\ll m_ec^2$ and neglect Klein-Nishina corrections.
Most of the radiation emitted by the neutron star has the $\perp$ polarization.

The energy loss of the electron due to scattering is given by
\beq
    \dE_e=-\int d\Omega \int dE\, (1-\beta\mu)\,\sigma(\Ec) 
                \,\frac{I(\nn,E)}{E}\,\left(\overline{E^\prime}-E\right),
\eeq 
where $\nn$ is the unit vector describing the photon direction in solid 
angle $d\Omega$, and $\overline{E^\prime}=\gamma\Ec$ is the mean expectation
for the photon energy after scattering. This gives,
\beq
\label{eq:dE_e}
   \dE_e=-\int d\Omega \int dE\, (1-\beta\mu) \left[\gamma^2(1-\beta\mu)-1\right]
        I(\nn,E)\,\sigma(\Ec)\,dE.
\eeq
In the simplest case of Thomson scattering of isotropic radiation, 
averaging over random $\mu$ gives the standard result 
$\dE_e=-(4/3)\sT\,c\,U\,\gamma^2\beta^2$, where $U$ is the energy density 
of radiation. In our case, $\sigma<\sT$, and the radiation field is not isotropic;
far from the star it is better approximated as a central beam.

Using \Eq~(\ref{eq:dE_e}) one can show that non-resonant scattering makes 
a small contribution to the radiative drag compared with resonant scattering,
and hence its inclusion in the calculation does not significantly change the 
outflow solution shown in Figure~9. Consider first the non-relativistic zone 
$p_+<1$. An upper bound on the non-resonant $\dE_e$ is obtained if we 
substitute into \Eq~(\ref{eq:dE_e}) $\sigma(\Ec)=\sT$ and $\mu=0$. 
This gives,
\beq
    \dE_e=\sT c\, p^2\,U.
\eeq
The drag coefficient due to non-resonant scattering is defined similar to 
\Eq~(\ref{eq:D}). Using $U\approx \Lth/4\pi r^2 c$ and 
$dp/dt=\dE_e/\beta m_ec^2$, one obtains
\beq
\label{eq:D1}
   \D\approx \frac{\sT\Lth\gamma}{4\pi\,r\,m_ec^3}
       \approx 0.2\, L_{\rm th,35}\,r_6^{-1}\,\gamma\ll 1.
\eeq  
In the relativistic zone $p_+>1$, the upper bound given by \Eq~(\ref{eq:D1}) 
increases proportionally to $\gamma$ and becomes useless, because it does
not take into account the strong reduction of the scattering cross section below 
$\sT$. In this zone, the outflow adjusts so that it can resonantly scatter photons 
with $E\sim 7kT$ and $\mu\sim -0.5$ (photons flowing from the equatorial 
reflector). This implies that the main targets for non-resonant scattering 
(photons flowing from the star with $E\sim 3kT$ and $\mu>0$) have energies
well below the resonance energy, $\Ec\sim (0.1-0.2)\hbar\omega_B$.
Then the scattering cross section is strongly reduced below $\sT$
according to \Eq~(\ref{eq:sig_nonres}). When this reduction is taken into
account, one obtains $\D<1$.

\end{appendix}



\end{document}